\documentclass[twocolumn,showpacs,preprintnumbers,amsmath,amssymb,floatfix]{revtex4}

\usepackage{dcolumn}
\usepackage{bm}

\usepackage{epic}
\usepackage{eepic}
\usepackage{graphicx}

\newcommand{\ket}[1]{|#1\rangle}
\newcommand{\bra}[1]{\langle #1|}
\renewcommand{\vec}[1]{{\bf{#1}}}
\newcommand{\uvec}[1]{{\bf{\hat #1}}}

\begin{document}

\title{Diffraction of ultra-cold fermions by quantized light
fields: Standing versus traveling waves}

\author{D. Meiser}
\author{C. P. Search}
\author{P. Meystre}
\affiliation{Optical Sciences Center, The University of Arizona,
Tucson, AZ 85721}

\begin{abstract}
We study the diffraction of quantum degenerate fermionic atoms off of
quantized light fields in an optical cavity. We compare the case
of a linear cavity with standing wave modes to that of a ring
cavity with two counter-propagating traveling wave modes. 
It is found that the dynamics of the atoms
strongly depends on the quantization procedure for the cavity
field. For standing waves, no correlations develop between the
cavity field and the atoms. Consequently, standing wave Fock
states yield the same results as a classical standing wave field
while coherent states give rise to a collapse and revivals in the
scattering of the atoms. In contrast, for
traveling waves the scattering results in quantum entanglement of 
the radiation field and the atoms. This
leads to a collapse and revival of the scattering
probability even for Fock states. The Pauli Exclusion Principle
manifests itself as an additional dephasing of the scattering
probability.
\end{abstract}


\maketitle

\section{\label{introduction}Introduction}

The past few decades have witnessed considerable progress in the
cooling of atomic vapors to extremely low temperatures,
culminating in the achievement of Bose-Einstein condensation in
dilute alkali gases
\cite{Cornell:BEC1995,Hulet:BEC1995,Ketterle:BEC1995}. More
recently, quantum degenerate Fermi gases with temperatures as low
as $0.01T_F$, where $T_F$ is the Fermi temperature, have been
achieved by several groups \cite{Regal:BEC_BCS_crossover,
Hadzibabic:ultracold_fermions,Strecker:fermi_gas}. Throughout
these developments the interaction of light with atoms has been
central to the cooling, trapping, and imaging of atoms, as well as
in the coherent manipulation of their center-of-mass motion. For
example, the Bragg scattering of atomic matter waves by
off-resonant optical fields can be used to create linear atom
optical elements for use in atom interferometers
\cite{Burgbacher:beam_splitter_bosons}, and the interaction of
atomic condensates with light has led to the realization of
matter-wave superradiance \cite{Ketterle:superradiance} and of
matter-wave parametric amplifiers
\cite{Kozuma:matter_wave_amplification,Ketterle:Matter_wave_amplification,Law:matter_wave_amplification}.
In another application, the ability of optical fields to create
custom trapping potentials has permitted the study of condensed
matter problems such as e.g. the Mott-Insulator transition
\cite{Jaksch:BECinLattice,Bloch:MottInsulator1,Bloch:MottInsulator2}.
Although all experiments to date have involved classical optical
fields, there is considerable interest in carrying out future work
in high-Q optical cavities, where the quantum nature of the
electromagnetic field becomes important. Theoretical work along
these lines has so far been restricted to the case of bosonic
atoms, see e.g. Ref.  \cite{Moore:qu_optics_bec}, while the
diffraction of fermions by an optical field was discussed in Ref.
\cite{Lenz:kapitza_dirac}, but in an analysis restricted to the
case of classical fields. In this paper we extend this work to
discuss the diffraction of quantum-degenerate fermionic
matter-wave fields by quantized light fields.

We consider a zero-temperature beam of fermionic two-level atoms
traversing an optical cavity supporting an off-resonant standing
wave light field of momentum $q$. The atoms undergo virtual
transitions to their excited electronic state, resulting in a
center-of-mass momentum recoil of $2q$ . Alternatively, one can
view this process as diffraction of the atoms off of the intensity
grating formed by the cavity field.

The normal modes in terms of which the electromagnetic field is
quantized are determined by the boundary conditions of the cavity.
In a linear cavity with perfectly reflecting mirrors we have
standing wave mode functions. In a ring cavity, the light field
has to fulfill periodic boundary conditions and this results in
running-wave mode functions. Two counter-propagating traveling
wave modes of equal frequency can be  superposed to yield a
stationary standing wave field.

Under most circumstances it is a question of mathematical
convenience which mode functions are used for the description of
the field. Physically, however, the two cases are not the same and
for fields containing only a few photons, the two quantization
procedures yield different results. In particular, the difference
in atomic scattering produced in these two situations has been
discussed for single atoms diffracted by a coherent light field
\cite{Shore:stand_trav_wave}. It was shown to depend  critically
on the quantization procedure. This
difference can be understood in terms of which-way information for
the scattering process. For standing wave modes, the state of the
light field contains no information about the momentum transfer to
the atom. More specifically, the number of photons is a constant of
motion and as a result the equations of motion for the atomic center of
mass decouple from that of the light field. In the case 
of two counter-propagating traveling
wave modes however, the number of photons in each mode does change and
the change in the number of quanta is a
direct  measure of the momentum transfer to the atoms. In this
paper we extend those results to compare the diffraction of a
quantum-degenerate Fermi gas by these fields, both in the
Raman-Nath and Bragg regimes.

In the Raman-Nath regime, which is characteristic of situations
where the kinetic energy of the atoms can be neglected, 
the individual atomic dynamics for a standing wave light field
are formally identical to the case of a
classical light field \cite{Meystre:elements_of_quantum_optics}.
The atoms scatter into successive diffraction orders separated by
twice the photon momentum $q$, up to the point where
energy-momentum conservation becomes important and the Raman-Nath
approximation ceases to hold. The formal equivalence of the
scattering off of a standing wave field to the scattering
off of a classical light field is due to the fact that the equations
of motion for the atoms effectively decouple both from each other and from the
light field. This must be contrasted to the case
of running waves, where the number operators for the two modes are
not constants of motion. This leads to an infinite hierarchy of
coupled equations for the atomic and optical field operators, 
with higher-order
correlation functions playing a crucial role in the dynamics of
first-order atomic correlation functions. It is then necessary to
introduce some approximate truncation scheme, a procedure that we
discuss in detail and compare with exact numerical results for
small atom numbers.

In the Bragg regime, energy-momentum conservation reduces the
single-atom diffraction problem to a two-mode situation, the atoms
undergoing Bragg oscillations between their initial momentum
states, $p_i$, and final momentum state $p_f=p_i+2q$. The
character of these oscillations is the result of three separate
and independent effects which correspond to whether one uses
standing wave or traveling wave modes, whether the cavity field is
in a Fock state or in a coherent state, and the momentum spread of
the incident atomic beam.

This paper is organized as follows: After formulating the specific
model used in our analysis in section \ref{model} we discuss the
case of traveling-wave light quantization in section
\ref{trav_wave}. We develop approximate equations for first and
second-order correlation function appropriate for the  Raman-Nath
regime, and a Bloch vector picture useful to discuss Bragg
diffraction. Specifically, that picture yields a semiclassical
model that provides some intuitive understanding of the atomic
dynamics. The case of standing-wave quantization is discussed in
Section \ref{stand_wave}. Section \ref{discussion} gives a summary and
conclusion.

\section{\label{model}Model}

\begin{figure}
\includegraphics[width=0.40\columnwidth]{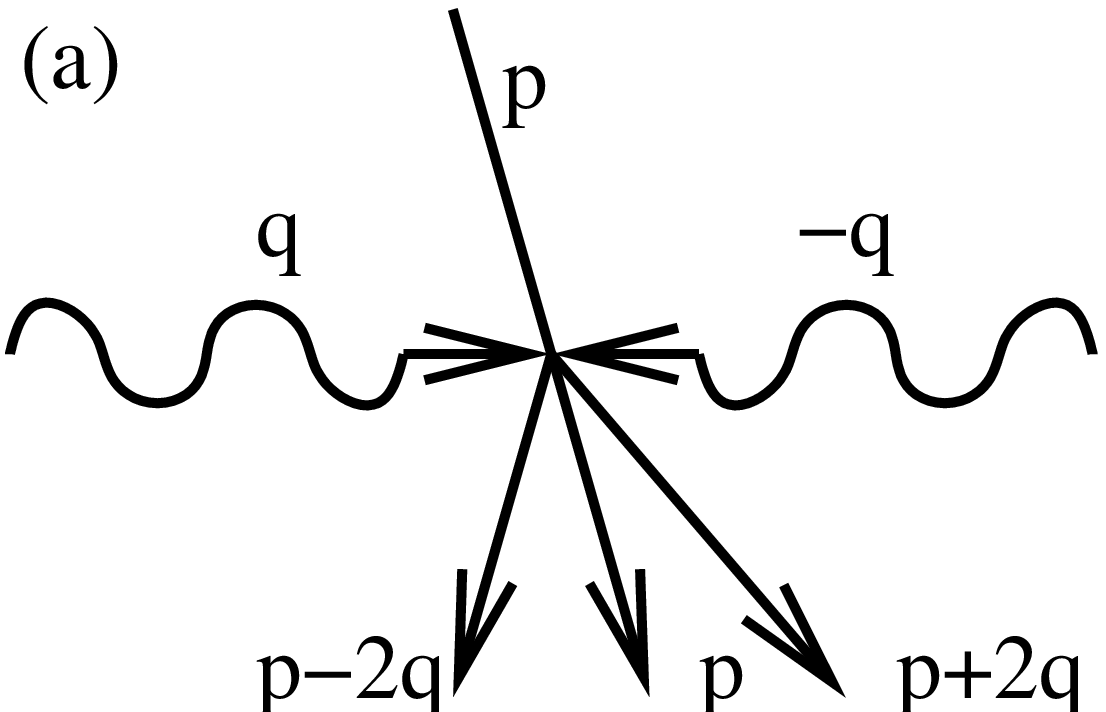}
\includegraphics[width=0.40\columnwidth]{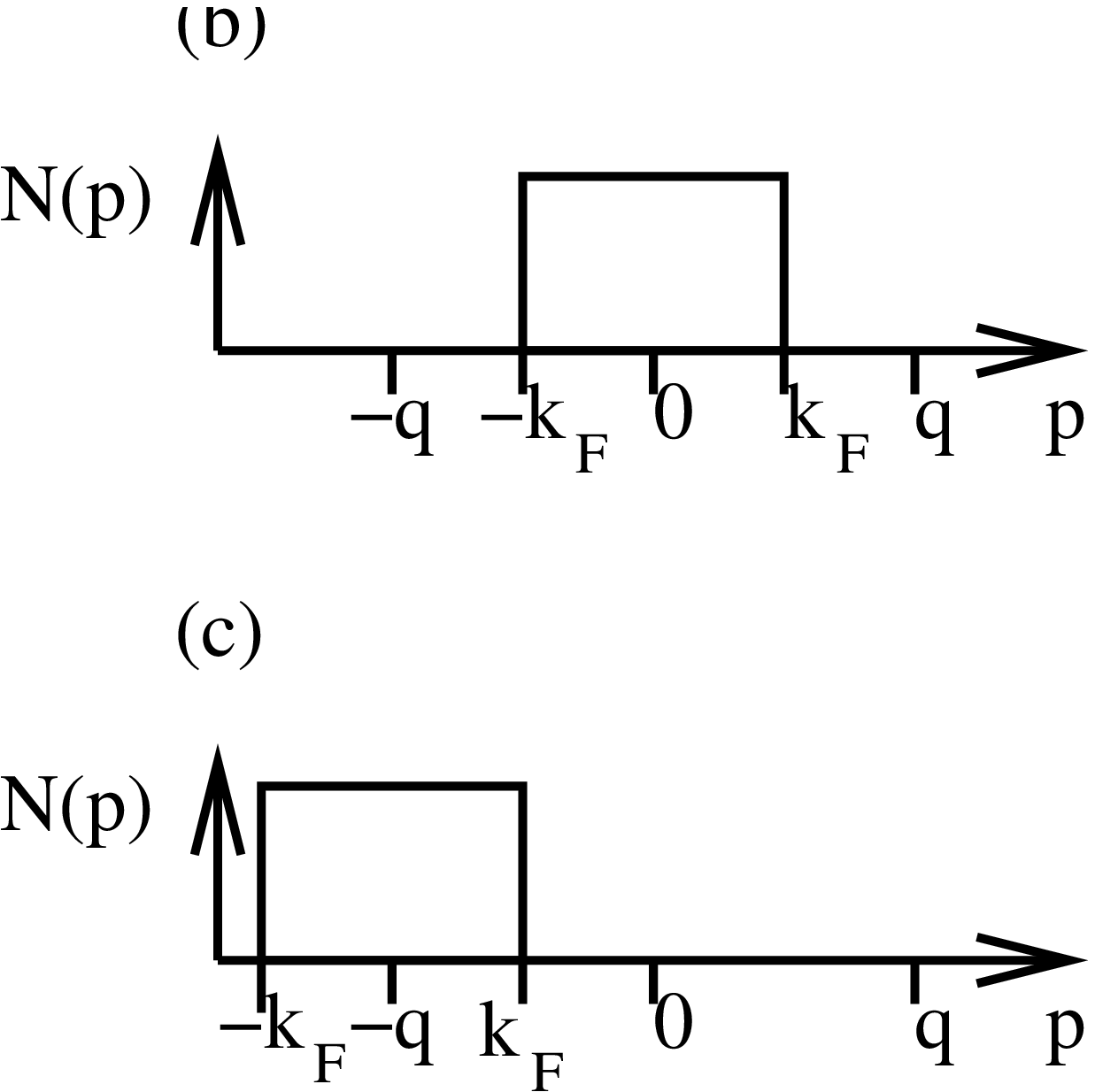}
\caption{(a) Schematic of a scattering of an atom of initial
momentum $p$ via two photon transitions with photons of momenta $q$
and $-q$. (b) Initial momentum distribution $N(p)$ of the atoms for
scattering in the Raman-Nath regime and (c) for the Bragg regime.}
\label{initialstate_fig}

\end{figure}

We consider an ultracold beam of identical two-level
fermionic atoms propagating across a high-$Q$ optical
cavity, see Fig. \ref{initialstate_fig}. Their initial momentum
distribution is a Fermi sea at $T=0$, but shifted in momentum space
by the  mean momentum
$\vec{p}$ and with Fermi momentum $k_F$ assumed to be much less
than $q$, the photon momentum. This is a realistic approximation,
since for a degenerate
Fermi gas of density $n\approx 10^{17} \text{m}^{-3}$ the Fermi
momentum is $k_F\approx 10^6 \text{m}^{-1}$ while for a photon of
wavelength $\lambda=500 \text{nm}$ one has a momentum of $q\approx
10^7 \text{m}^{-1}$. In the following we neglect atomic collisions --
a good approximation at low temperatures, since the $s$-wave scattering
length is zero for identical fermions -- as well as cavity losses. We
also assume that the optical frequency $\omega$ is sufficiently detuned
from the atomic transition frequency $\omega_0$ that the upper electronic
level can be adiabatically eliminated. Finally, we consider a situation
where that atomic momentum $m v_\perp$ transverse to the cavity field is
large enough so that it can be treated classically. Time, $t$,
can then be parameterized in terms of the transverse distance $x$
by $t=x/v_\perp$.

\section{\label{trav_wave} Running waves}

For running-wave quantization, the Hamiltonian describing our system
is $(\hbar = 1)$

\begin{eqnarray}
H_r &=&
\sum_k E_k c^\dagger_k c_k +
\omega(a_q^\dagger a_q + a_{-q}^\dagger a_{-q} + 1 )\nonumber\\
&+& \left ( g a_q^\dagger a_{-q}\sum_k c^\dagger_{k-q}c_{k+q} + h. c. \right ),
\label{basic_hamiltonian_rw}
\end{eqnarray}
where, $c_k$ and $c_k^\dagger$ are the annihilation and creation
operators for a fermionic atom of momentum $k$, $a_q$ and
$a_q^\dagger$ are the annihilation and creation operators for a
photon of momentum $q$, $E_k=k^2/2M$ is the kinetic energy of an
atom of momentum $k$, $g=\Omega_R^2/\Delta$ is the coupling energy
of the atoms and the light field, $\Omega_R$ is the vacuum Rabi
frequency, and $\Delta=\omega-\omega_0$ is the atom-light
detuning.

The initial state of the atoms-field system is
\begin{equation}
\ket{\psi(0)}_{\rm rw}=
\ket{\phi_q}\ket{\phi_{-q}}
\prod_{|k|<k_F}c_{k}^\dagger \ket{0}.
\label{initial_rn_rw}
\end{equation}
where the field states $|\phi_{\pm q} \rangle$ are taken
to be either Fock states $|N_{\pm q}\rangle$ or coherent
states $|\alpha_{\pm q}\rangle$.

\subsection{Raman-Nath regime}

The Raman-Nath regime of atomic diffraction is characteristic of situations
where the kinetic energy of the atoms plays a negligible role in comparison with the interaction energy, i.e. $E_{2q}\ll g\sqrt{N_q N_{-q}}$, where the recoil energy $E_{2q}=2q^2/M$ is a measure for the typical kinetic energies involved. 
In practice, this amounts to assuming that the atoms have an infinite mass, and as such,
neglects the effects of the quadratic dispersion relation of the atoms.

The most straightforward way to solve this problem proceeds by integrating
the Schr\"odinger equation corresponding to
Hamiltonian (\ref{basic_hamiltonian_rw}) for the initial conditions
(\ref{initial_rn_rw}), From which we can obtain the probability
\begin{equation}
P_{p}(t)=\bra{\psi(t)}c_{p}^\dagger c_{p}\ket{\psi(t)}.
\end{equation}
for an atom being scattered to a state of momentum $p$.
However, the dimension of the Hilbert space grows exponentially as
${\rm Dim}_{\rm Raman-Nath}=(2n_d+1)^{N_{a}}(N_{p}+1)$
where $n_d$ is the number of diffraction orders considered, $N_a$ is the
number of atoms, and $N_p$ the total number of photons. Hence, a direct
integration of the Schr\"odinger
equation is only possible for rather small atom and photon numbers.

\begin{figure*}
\includegraphics[width=0.45\textwidth]{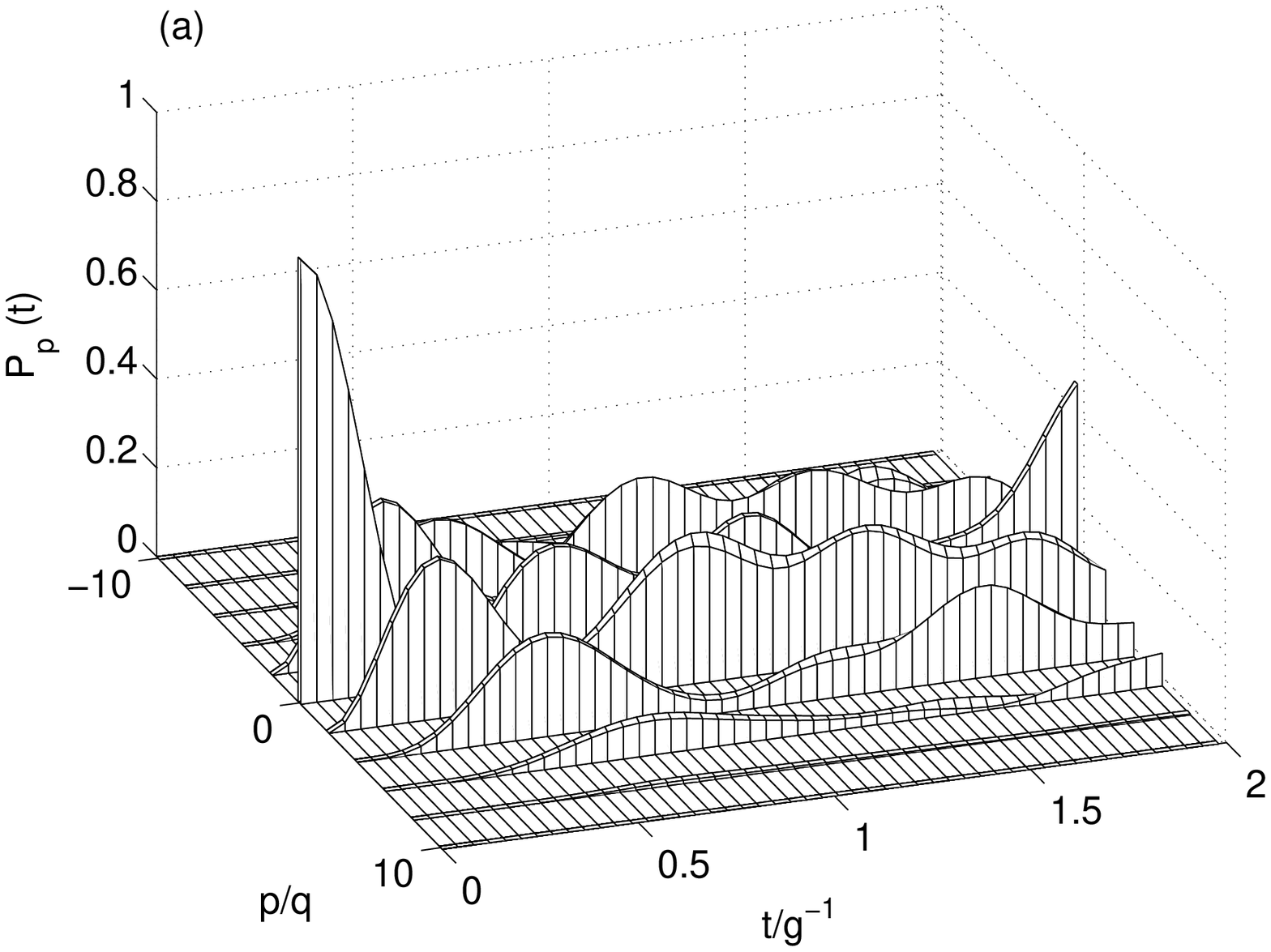}
\includegraphics[width=0.45\textwidth]{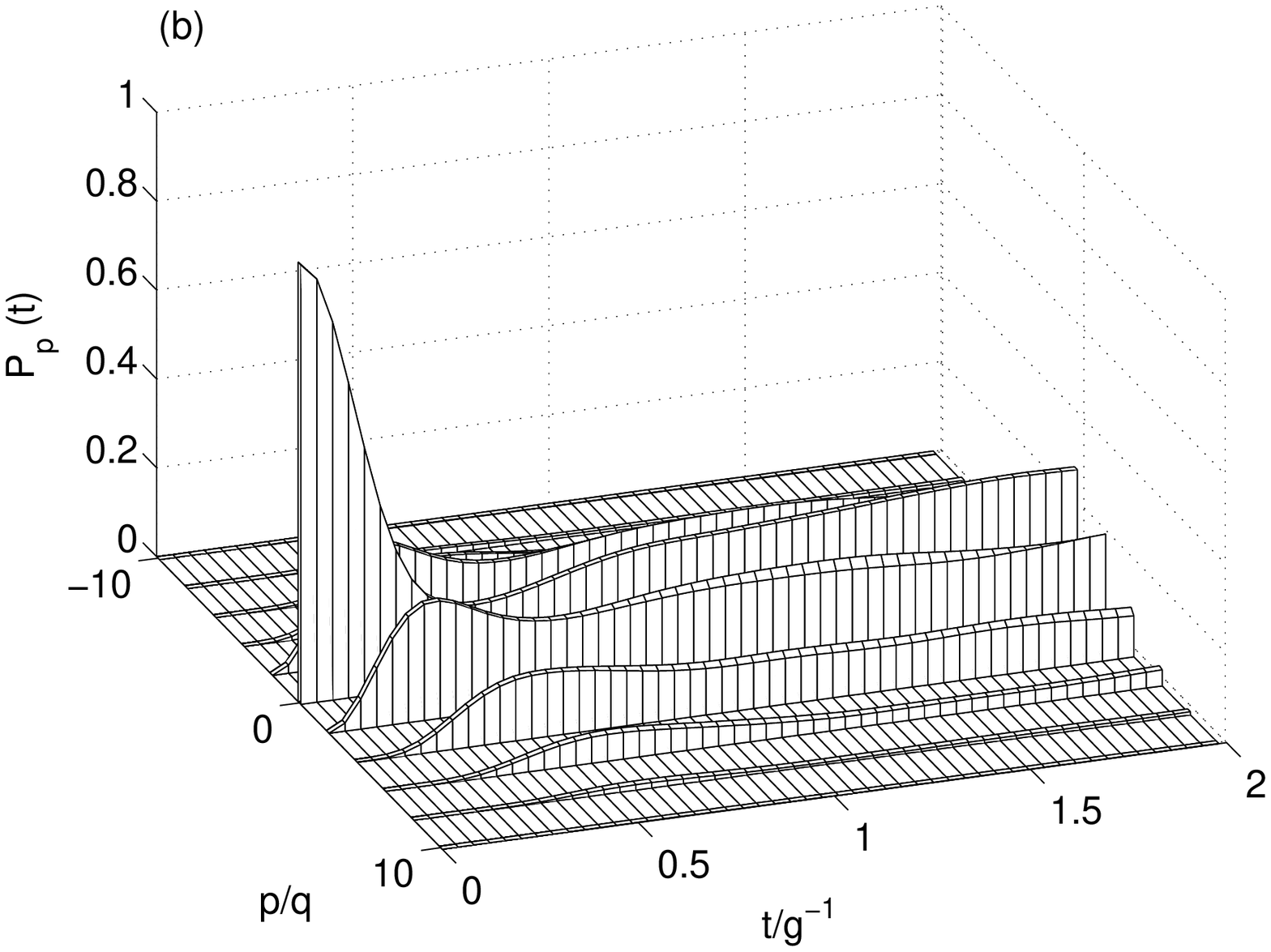}
\includegraphics[width=0.45\textwidth]{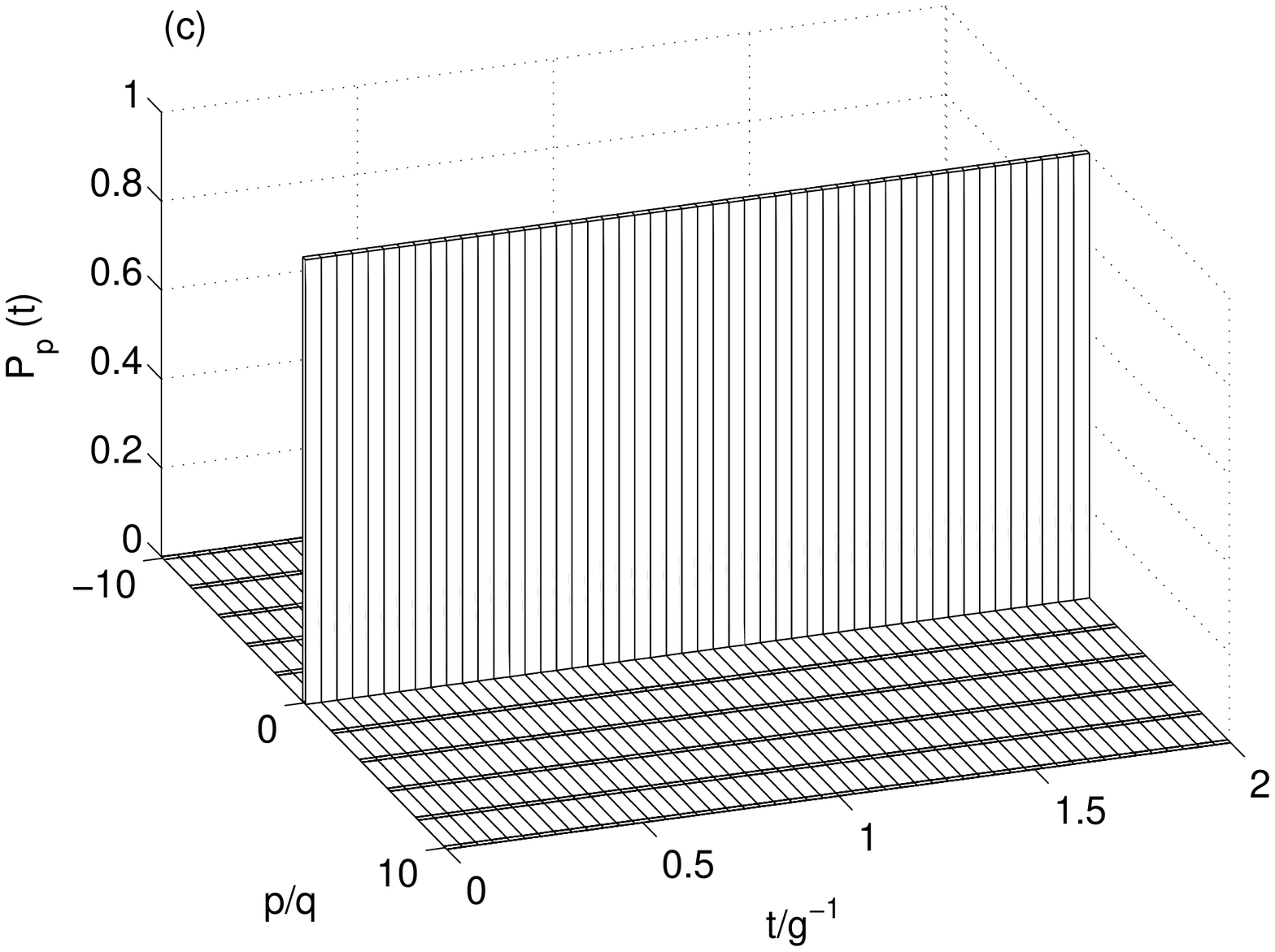}
\includegraphics[width=0.45\textwidth]{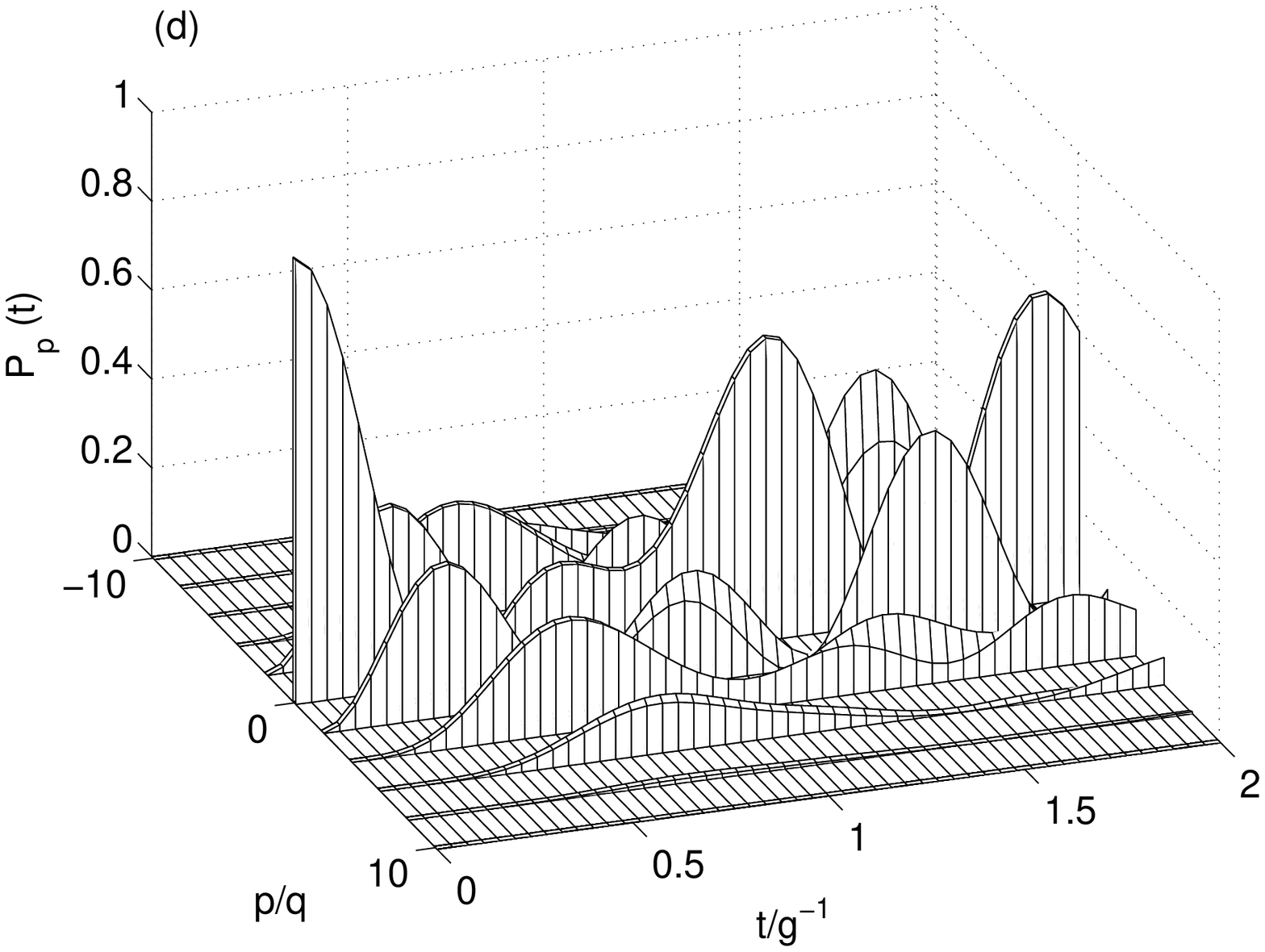}
\includegraphics[width=0.45\textwidth]{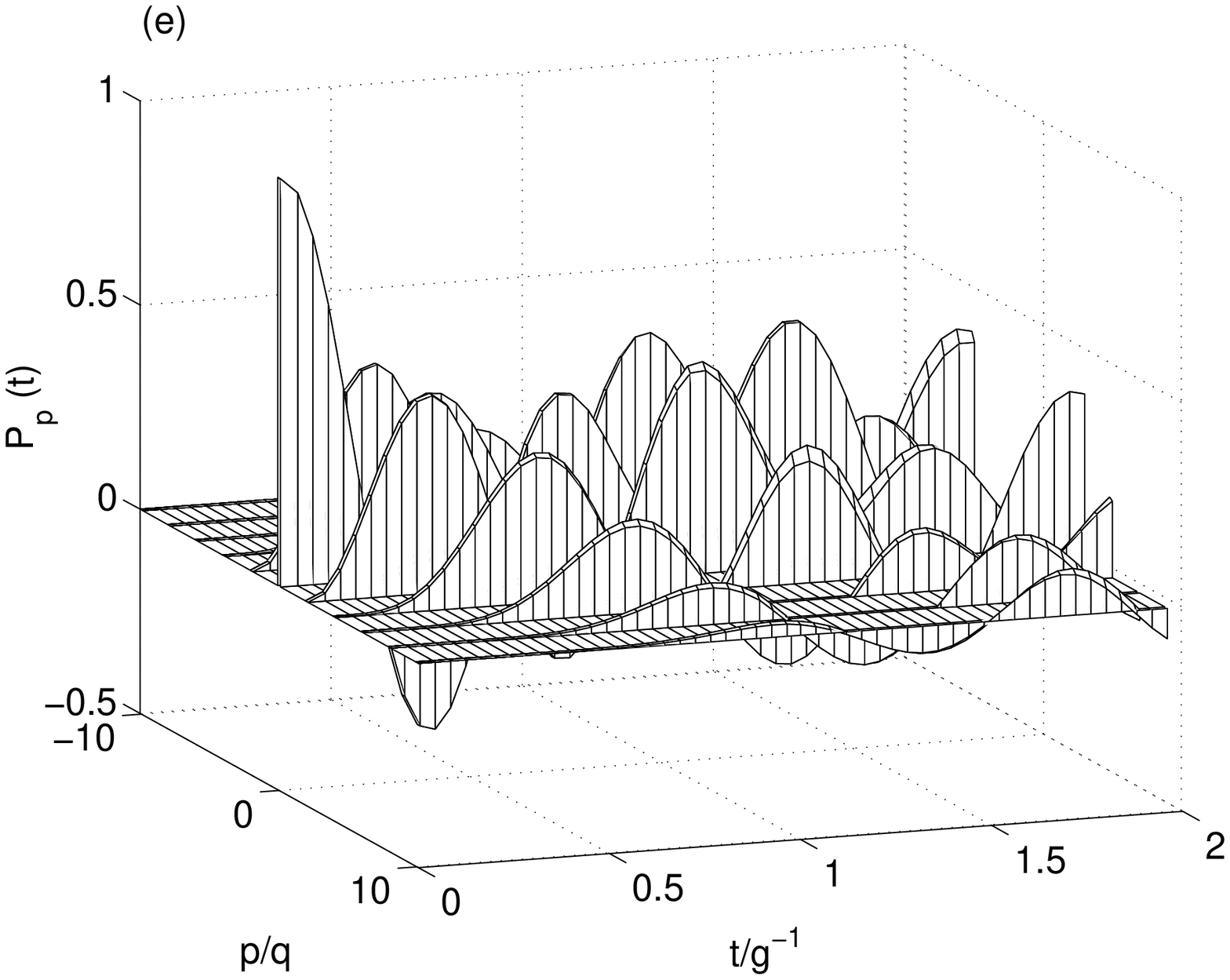}
\includegraphics[width=0.45\textwidth]{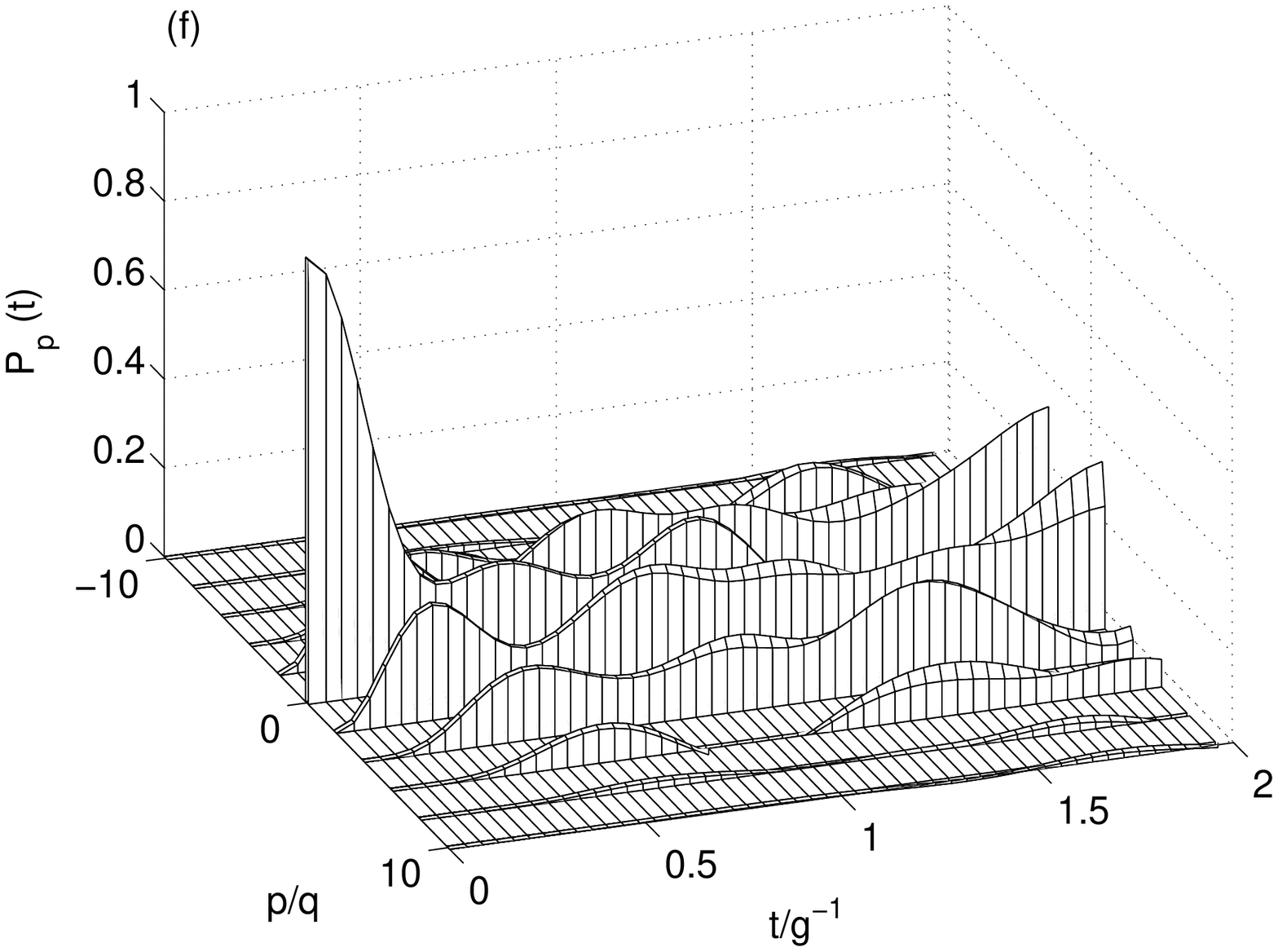}
\caption{Scattering probability $P_{p}(t)$ for two atoms scattering
off of a running-wave
light field in the Raman-Nath regime.
Figures (a,c,e) are for a Fock state of the light field and figures (b,d,f)
for a coherent state. Figures (a,b) show the exact solution of the
Schr\"odinger equation, figures (c,d) for the first-order equations
and figures (e,f) show the results for the second-order equations.
In all the calculations the recoil energy is $E_{2q}=g$ and the Fermi
momentum is $k_F=0.1$. Time is in
units of $g^{-1}$ and momentum in units of the photon momentum $q$.}
\label{fig_rn_rw}
\end{figure*}

Figure \ref{fig_rn_rw}(a) shows the result of an exact solution of
the Schr{\"o}dinger equation for $N_a=2$ and the light field in a
Fock state with three photons per mode initially. In this example, the
recoil energy is $E_{2q}=g$ and the initial momentum of
the atoms is $p_i=\pm 0.1q$. Such a high recoil energy was chosen to
limit the number of diffraction orders
that are significantly populated before energy-momentum
conservation inhibits further diffraction, i.e. before exiting the
Raman-Nath regime.

The resulting dynamics resembles qualitatively the single-atom
case, see e.g. \cite{Meystre:elements_of_quantum_optics}. For
short times the probability for finding an atom in the $m$th order
mode is well described by $\sim J_m^2(2gt)$ where $J_m$ is the
$m$-th Bessel function. For longer interaction times, higher
scattering orders are suppressed due to energy-momentum
conservation, as expected. We note that since the difference in
kinetic energies of the two atoms is small compared to all other
relevant energies, we do not observe any effect of ``inhomogeneous
broadening.''

For comparison, the results for initial coherent states with mean
photon numbers $\overline{N}_q=\overline{N}_{-q}=3$ are shown in
Fig. \ref{fig_rn_rw}(b), the atomic parameters being the same as
before. We now observe a decay of the oscillations of the
scattering probabilities after a time $t\sim (2\pi/g)
(\overline{N}_q\overline{N}_{-q})^{-1/2}$, which corresponds to a
complete dephasing of the contributions of the different photon numbers
to the diffraction pattern.

In order to proceed past the few-atom problem, we now concentrate
on single-particle properties, introducing a BBGKY-type truncation
scheme to factorize higher-order correlation functions of the
matter-wave field. From the Hamiltonian
(\ref{basic_hamiltonian_rw}), the equations of motion for the
atomic first order correlations $\langle c_{k_1}^\dagger c_{k_2}
\rangle$ and $\langle a_{q_1}^\dagger a_{q_2}\rangle $,
\begin{widetext}
\begin{equation} i\frac{d}{dt}\langle c_{k_1}^\dagger c_{k_2}\rangle =
 (E_{k_2}-E_{k_1})\langle c_{k_1}^\dagger
c_{k_2} \rangle
+g\left \langle a_{-q}^\dagger a_q\left(c_{k_1}^\dagger c_{k_2-2q} -
c_{k_1+2q}^\dagger
c_{k_2}\right) \right \rangle
+g \left \langle a_q^\dagger a_{-q}\left(c_{k_1}^\dagger c_{k_2+2q} -
c_{k_1-2q}^\dagger c_{k_2}\right)\right \rangle,
\label{eqnofmotion_rw_rn_atoms}
\end{equation}
\begin{equation}
i\frac{d}{dt}\langle a_{q_1}^\dagger
a_{q_2} \rangle =
 g \sum_k \left\langle \delta_{q_2
,q}a_{q_1}^\dagger
a_{-q}c_{k-q}^\dagger c_{k+q}+\delta_{q_2 ,-q}
a_{q_1}^\dagger a_q c_{k+q}^\dagger c_{k-q}
+\delta_{q_1,-q}a_q^\dagger a_{q_2} c_{k-q}^\dagger
c_{k+q}
+\delta_{q_1,q}a_{-q}^\dagger a_{q_2} c_{k+q}^\dagger
c_{k-q}\right \rangle,
\label{equnofmotion_rw_rn_light}
\end{equation}
\end{widetext}
where the $\delta$'s are Kronecker-deltas.

The simplest factorization scheme consists in merely factorizing
second-order correlation functions of the type $\langle
a_{q_1}^\dagger a_{q_2}c_{k_1}^\dagger c_{k_2}\rangle$ that appear
on the right-hand side of these equations into products of
first-order correlation functions, for instance, $\langle
a_{q_1}^\dagger a_{q_2}c_{k_1}^\dagger c_{k_2}\rangle
\approx\langle a_{q_1}^\dagger a_{q_2}\rangle \langle
c_{k_1}^\dagger c_{k_2}\rangle$ . In doing so we neglect
correlations that may build up between the atoms and the light
field as well as higher-order correlations of both the atoms and
the light field. This corresponds to a truncation of the
BBGKY-type hierarchy of the equations of motion for the
higher-order moments of the particle-hole operators after the
first order, see e.g. \cite{Kerson_Huang}. This reduces the
infinite hierarchy of equations
(\ref{eqnofmotion_rw_rn_atoms}-\ref{equnofmotion_rw_rn_light}) to
a closed set of $c$-number equations that grows only quadratically
with the number of momentum states that have to be taken into
account.

The result of the numerical integration of these equations of
motion, using the same parameters as previously for ease of
comparison, is shown in Fig.\ref{fig_rn_rw}(c,d) for the cases of
a Fock state and a coherent state of the field, respectively.

An obvious weakness of the simple truncation scheme is that it
predicts the absence of scattering for the case of Fock states, in
stark contrast to the exact solution. This follows from the
absence of initial coherence in either the light field or the
atoms, leading to the scattering term in
(\ref{eqnofmotion_rw_rn_atoms}) being identically zero. Stated
differently, the reason for the absence of diffraction is that the
phase of a Fock state is completely undetermined, hence there is
no established relative phase between the two counter-propagating
fields, and no light intensity grating. Since in this
factorization scheme the atom is effectively assumed to probe only
first-order moments of the light field, that is, its intensity
pattern, diffraction is absent at this level of approximation.

The situation is different for a coherent state light field. In
this case, there is a well-established phase relationship between
the two modes. This results in an intensity grating from which the
atoms can be diffracted. As time goes on, this results in the
generation of atomic coherence, $\langle c_{k_1}^\dagger
c_{k_2}\rangle\neq 0$, and the resulting density grating formed by
the atoms acts back on the light field. In some loose sense, the
lowest order factorization scheme consists in treating the system
classically since it neglects all quantum fluctuations in the atomic
and optical fields. It is not surprising that this approach should fail
for a very non-classical field state such as a Fock state, and be
much better for a quasi-classical field. Note however that while
for short enough times the scattering closely resembles the exact
results, this is no longer the case for long times, a consequence
of the build-up of quantum correlations between the optical and
matter-wave fields. Even after one oscillation differences arise.

We note that for our specific initial
conditions, the fully factorized equations for the light field can
be trivially integrated, showing that the first-order moments of
the light field are constants of motion. Inserting these constants
in the atomic equations of motion shows that at this level, the
scattering becomes formally equivalent to the scattering of atoms
by a classical standing wave light field with intensity 
$g\langle a_q^\dagger a_{-q}\rangle$. This is further
discussed in the following section.

The equations of motion
(\ref{eqnofmotion_rw_rn_atoms}-\ref{equnofmotion_rw_rn_light}),
suggest that an improved factorization scheme would retain the
lowest order correlations between light field and atoms. In order
to do so, we supplement the equations of motion for $\langle
c_{k_1}^\dagger c_{k_2}\rangle$ and $\langle a_{q_1}^\dagger
a_{q_2}\rangle$ by equations of motion for the cross-correlations
$\langle a_{q_1}^\dagger a_{q_2}c_{k_1}^\dagger c_{k_2}\rangle$ and
the second order correlations of the lightfield and the atoms.
This should remedy the major flaw of the first-order calculation,
namely its inability to predict atomic scattering for a light
field in a Fock state.

The equations for the lowest-order atom-field correlation
functions involve third-order correlations of the form $\langle
a_{q_1}^\dagger a_{q_2}a_{q_3}^\dagger a_{q_4}c_{k_1}^\dagger
c_{k_2}\rangle$ and $\langle a_{q_1}^\dagger
a_{q_2}c_{k_1}^\dagger c_{k_2}c_{k_3}^\dagger  c_{k_4}\rangle$. We
truncate the resulting hierarchy of equations of motion by
introducing the factorization scheme
\begin{widetext}
\begin{equation}
\label{thirdorder_factorization}
\langle a_{q_1}^\dagger a_{q_2}a_{q_3}^\dagger
a_{q_4}c_{k_1}^\dagger c_{k_2}\rangle \simeq
\langle a_{q_1}^\dagger a_{q_2}\rangle
\langle a_{q_3}^\dagger a_{q_4}c_{k_1}^\dagger c_{k_2}\rangle
+\langle a_{q_1}^\dagger a_{q_2}a_{q_3}^\dagger
a_{q_4}\rangle\langle c_{k_1}^\dagger c_{k_2}\rangle
+\langle a_{q_1}^\dagger a_{q_2}c_{k_1}^\dagger
c_{k_2}\rangle\langle a_{q_3}^\dagger a_{q_4}\rangle
-2\langle a_{q_1}^\dagger a_{q_2}\rangle\langle a_{q_3}^\dagger
a_{q_4}\rangle\langle c_{k_1}^\dagger c_{k_2}\rangle
\end{equation}
\end{widetext}
and similarly for $\langle a_{q_1}^\dagger a_{q_2}c_{k_1}^\dagger
c_{k_2}c_{k_3}^\dagger c_{k_4}\rangle$ with $a_q$'s replaced by
$c_k$'s and vice versa \cite{Anglin:BEC_beyond_MFT}.
The last term of this equation accounts for the case where all
first-order correlation functions are uncorrelated.

\begin{figure}
\includegraphics[width=\columnwidth]{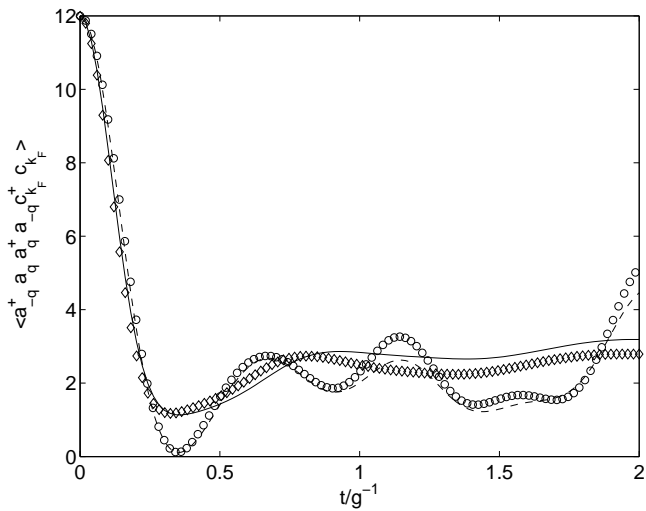}
\caption{Expectation value of $a_{-q}^\dagger a_{q}a_{q}^\dagger
a_{-q}c_{k_F}^\dagger c_{k_F}$ as calculated from the numerical
solution of the Schr\"odinger equation for a Fock state and for a
coherent state of the light field with the same parameters that
were used in Fig. \ref{fig_rn_rw}. $\circ$ and $\diamond$ show
the exact unfactorized value for Fock state and coherent state
respectively; the broken line ($-\hspace{0.1cm} -$) and the solid
line (---) show the corresponding values for the Fock state and
coherent state as obtained with the factorization scheme
(\ref{thirdorder_factorization}).}
\label{factorizationestimate}
\end{figure}

\begin{figure}
\includegraphics[width=\columnwidth]{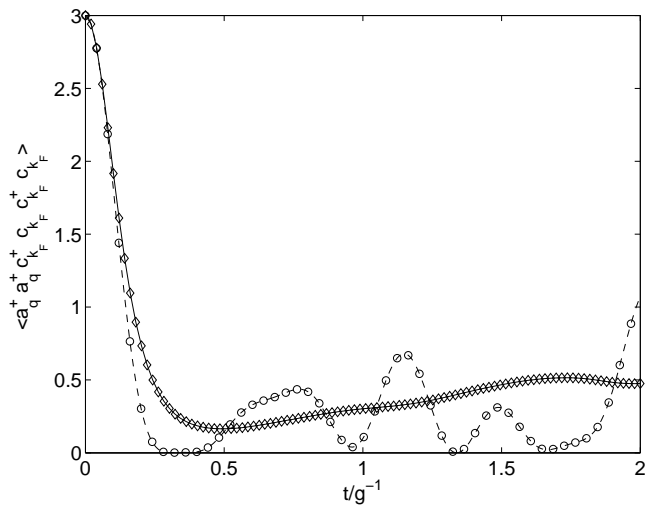}
\caption{Expectation value of $a_q^\dagger a_q c_{k_F}^\dagger c_{k_F}
c_{k_F}^\dagger c_{k_F}$ as calculated from the numerical solution of
the Schr\"odinger equation for a Fock state and for a coherent
state of the light field with the same parameters that were
used above. $\circ$ and $\diamond$ show the exact unfactorized
value for Fock state and coherent state respectively. The
broken line ($-\hspace{0.1cm} -$) and the solid line
(---) show the values obtained from the factorization scheme
(\ref{thirdorder_factorization}) for the two cases.}
\label{thirdorderfactorizationaacccc}
\end{figure}

We estimate the accuracy of the factorization scheme by
calculating $\langle a_{q_1}^\dagger a_{q_2}a_{q_3}^\dagger
a_{q_4}c_{k_1}^\dagger c_{k_2}\rangle$  and $\langle
a_{q_1}^\dagger  a_{q_2} c_{k_1}^\dagger c_{k_2} c_{k_3}^\dagger
c_{k_4}\rangle$ as well as their respective factorized values, Eq.
(\ref{thirdorder_factorization}) using the exact solution of the
Schr\"odinger equation. As an example Fig.
\ref{factorizationestimate} shows the results for $\langle
a_{-q}^\dagger a_{q}a_{q}^\dagger a_{-q}c_{k_F}^\dagger
c_{k_F}\rangle$ and Fig. \ref{thirdorderfactorizationaacccc} shows
the results for $\langle a_q^\dagger a_q c_{k_F}^\dagger c_{k_F}
c_{k_F}^\dagger c_{k_F} \rangle$ for the parameters of Fig.
\ref{fig_rn_rw}. This shows that the factorization scheme
reproduces at least qualitatively the main features of the
third-order correlation functions for both coherent states and
Fock states. 

Despite its apparent success, we must keep in mind that this
factorization scheme suffers from two major flaws. First, the
small deviations of the factorized values from the exact values
will accumulate in the course of time, leading to increasing
discrepancies between the approximate and exact results. More
critical perhaps, this scheme violates important relations that 
the exact operators have to obey. For example, $a_q^\dagger a_q
c_{k_F}^\dagger c_{k_F} c_{k_F}^\dagger c_{k_F}$ is a positive
self-adjoint operator, with positive and real expectation values,
but the factorized approximation can take on negative values.
These flaws eventually result in non-physical behavior such as
illustrated in Fig. \ref{fig_rn_rw}(e,f) where the probabilities
$\langle c_k^\dagger c_k\rangle$ take on negative values. We have
not found a factorization scheme that avoids non-physical behavior
of that kind at all times, and conjecture that the factorization
of higher-order moments in lower-order moments necessarily leads
to such inconsistencies.

The results of the factorization scheme
(\ref{thirdorder_factorization}) are shown in Fig.
\ref{fig_rn_rw}(e) for a Fock state and in Fig. \ref{fig_rn_rw}(f)
for a coherent state of the light field. While a Fock state now
leads to atomic diffraction, as should be the case, it is
characterized by non-physical negative probabilities already for
short times. This is clear evidence that higher order correlations
play an essential role. This is in contrast with the situation for
a coherent state, where we achieve good agreement with the exact
results for times up to  $\sim (2\pi/g)(\overline{N}_q
\overline{N}_{-q})^{-1/2} $, indicating that the first and
second-order correlations are the most important.

\begin{figure}
\includegraphics[width=\columnwidth]{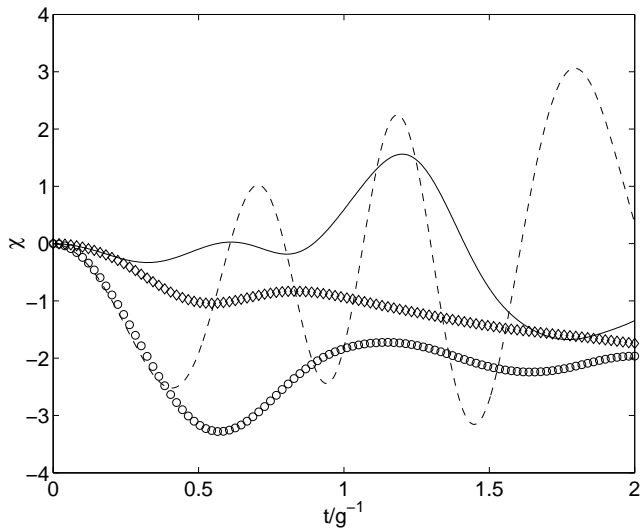}
\caption{Cross-correlation $\chi(t)$ between light field and atoms
in the Raman-Nath regime as obtained from the exact solution of
the Schr\"odinger equation for a Fock state ($\circ$) and a
coherent state of the light field ($\diamond$). Also shown are the
values obtained from the factorization scheme
(\ref{thirdorder_factorization}) ($-\hspace{0.1cm} -$) for a
Fock-state light field and (---) for a  coherent state. Same
parameters as in Fig. 2.} \label{crosscorr_fig}
\end{figure}

A quantitative measure of the degree of entanglement between the
atoms and the light field is given by the second-order
cross-correlation
\begin{equation}
\chi(t) = \sum_k\left(
\langle a_q^\dagger a_{-q} c_{k-q}^\dagger c_{k+q}\rangle -
\langle a_q^\dagger a_{-q}\rangle\langle c_{k-q}^\dagger c_{k+q}\rangle\right),
\end{equation}
which is equal to zero in the absence of entanglement. Figure
\ref{crosscorr_fig} shows $\chi(t)$ for both a Fock state and for
a coherent state light field, for the parameters of Fig.\ref{fig_rn_rw}.
Because the light field and atoms are initially uncorrelated we
have $\chi(t=0)=0$ but cross-correlations then build up to
become of the order of
$(\overline{N}_q\overline{N}_{-q})^{1/2}N_{a}$. The figure also
shows the result of the factorization ansatz
(\ref{thirdorder_factorization}), showing the good agreement with
the exact result for short enough times.

\subsection{Bragg regime}

In the Bragg regime, energy-momentum conservation restricts the
scattering of the atoms to two diffraction orders, an initial mode
of transverse momentum $p_i\approx -q$ and a final mode of
momentum $p_f=p_i+2q\approx q$. Classically, the atoms are known
to undergo Pendell{\"o}sung oscillations between these two modes.
As such, the atoms can be thought of as two-state systems that are
conveniently described in terms of pseudo-spin operators
\begin{eqnarray}
S_k^z&=&\frac{1}{2}\left (c_{k+q}^\dagger c_{k+q} -
c_{k-q}^\dagger
c_{k-q}\right ) \nonumber \\
S_k^+ &=& (S_k^-)^\dagger = c_{k+q}^\dagger c_{k-q}.
\label{pseudospin}
\end{eqnarray}
Introducing further the Schwinger representation of the light
field by means of
\begin{eqnarray}
J^z&=&\frac12 \left (a_q^\dagger a_q - a_{-q}^\dagger a_{-q}\right)\
\nonumber,\\
J^+ &=& (J^-)^\dagger = a_q^\dagger a_{-q},
\end{eqnarray}
the Hamiltonian of the atoms-field system simplifies to
\begin{equation}
H=\sum_{k\in [-k_F,k_F]} \left (\delta\omega_k S_k^z + g J^+S_k^-
+ g J^-S_k^+ \right ). \label{bragg_hamiltonian_rw}
\end{equation}
In this representation, the eigenvalues $m$ of $J^z$ correspond to
the photon number difference $m=(1/2)(N_q-N_{-q})$ between the two
counter-propagating modes and the eigenvalues $j(j+1)$ of
$J^2={J^z}^2+1/2(J^+ J^- + J^- J^+)$ correspond to the total
number of photons, $j=1/2(N_q+N_{-q})$. Finally, $\delta\omega_k =
E_{k+q}-E_{k-q} = 2 kq/M$ is the frequency mismatch between
the two momentum states accessible to the atom with initial momentum
$k-q$.

When compared to the Raman-Nath case, the dimension of the
Hilbert-space is now reduced to ${\rm Dim}_{\rm Bragg} = (
N_{p}+1)2^{N_{a}}$, which allows us to consider larger atomic and
photon numbers. The initial state of the atoms is now a $T=0$
Fermi sea shifted by $-q$ in momentum.  In the following, we
evaluate the total number of atoms diffracted to states of
momentum near $+q$,

\begin{equation}
N_{sc}(t)=\sum_{k\in [-k_F,k_F]} P_{k+q}(t).
\end{equation}
for a light field initially in a Fock state and in a coherent state.

\subsubsection{Fock state}

\begin{figure*}
\includegraphics[width=0.48\textwidth]{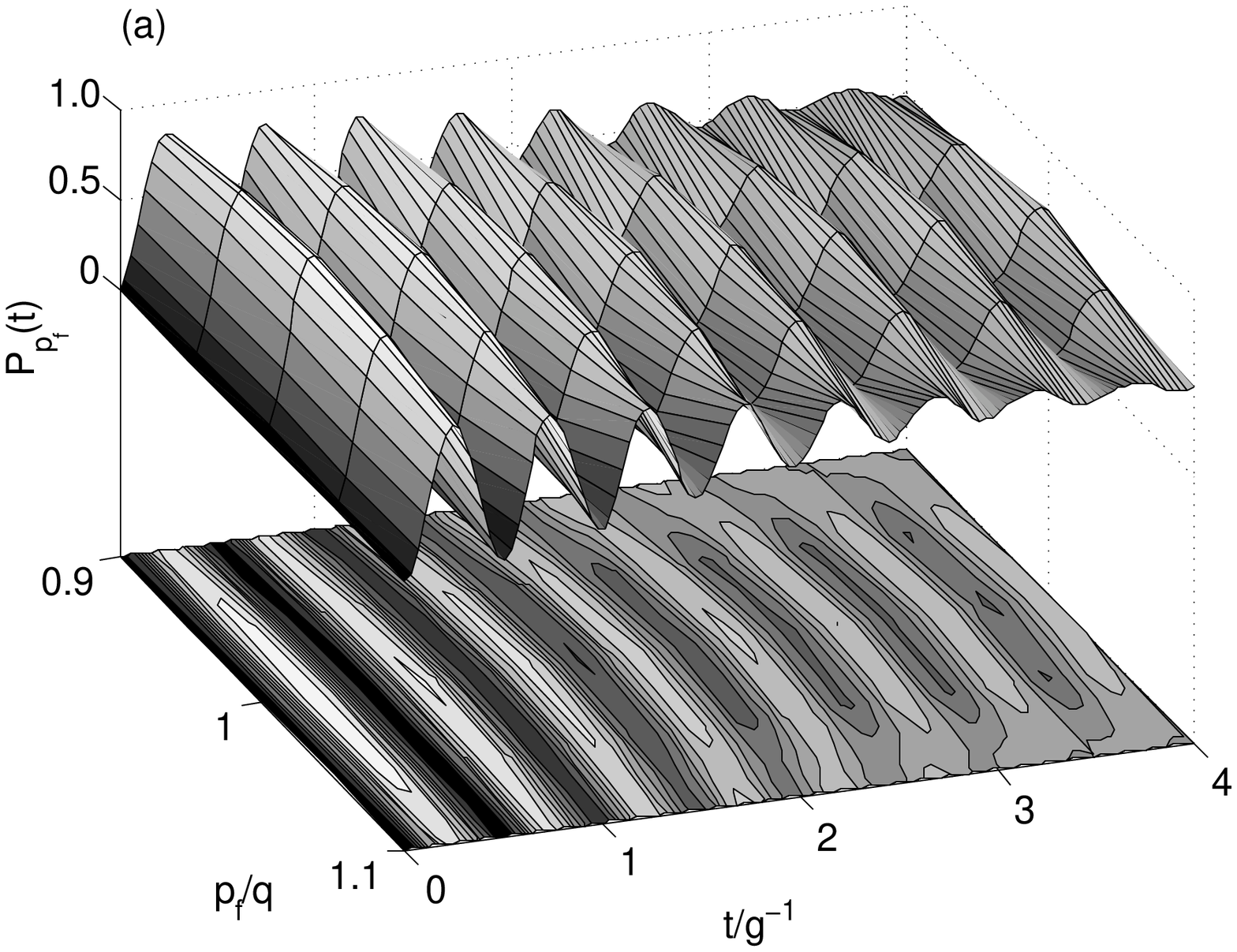}
\includegraphics[width=0.48\textwidth]{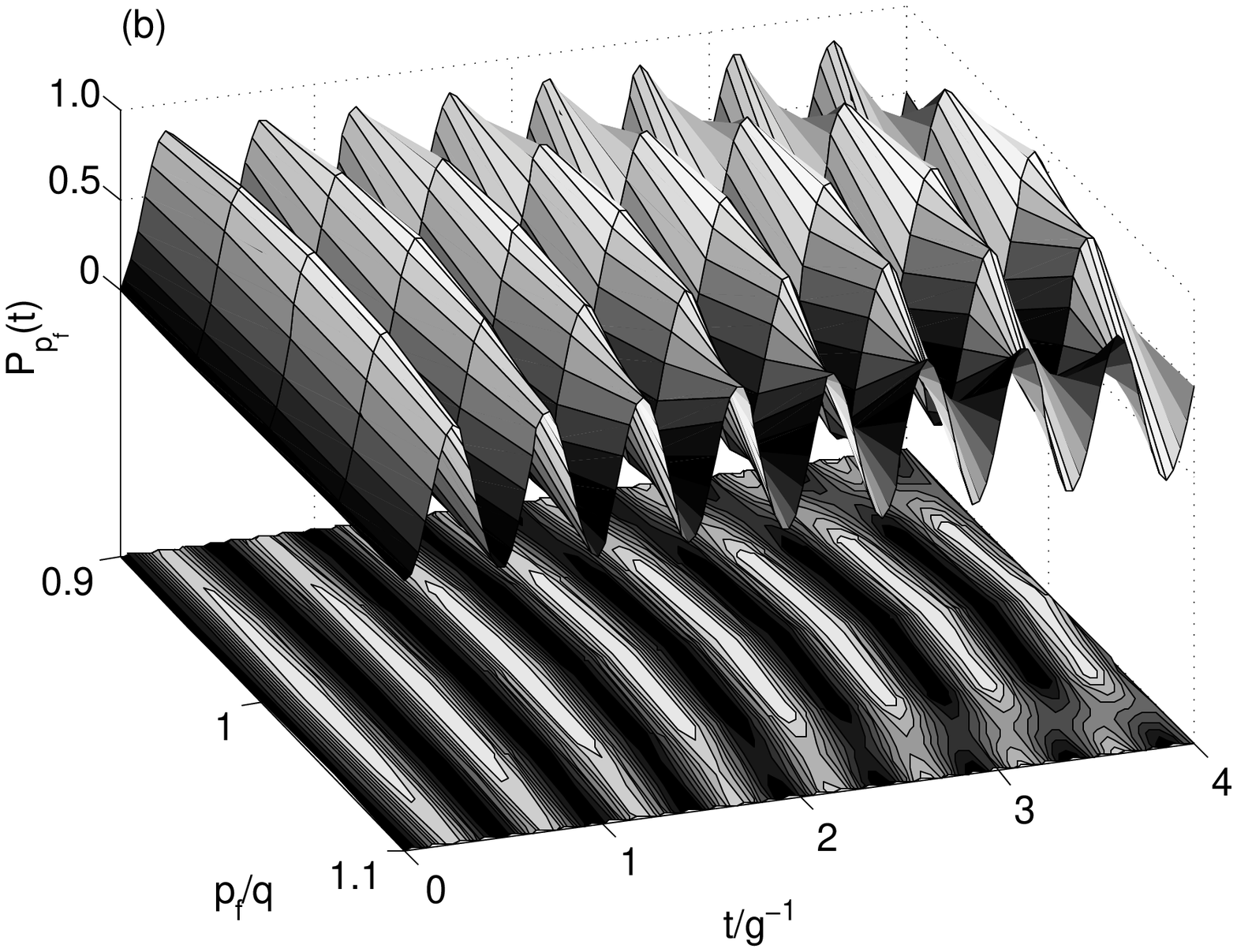}
\caption{Scattering probability
$P_{p_f}(t)$
for five fermions scattering off of a running-wave light field (a)
and a standing-wave light field (b) in the Bragg regime. In both
cases the light field is in a Fock state with $N_q=N_{-q}=6$ and
$N=12$ respectively, $E_{2q}=50g$ 
and $k_F=0.1q$. Time in units of $g^{-1}$ and momentum in units of $q$.}
\label{fig_rw_fock_pk}
\label{fig_sw_fock_pk}
\end{figure*}

\begin{figure}
\includegraphics[width=\columnwidth]{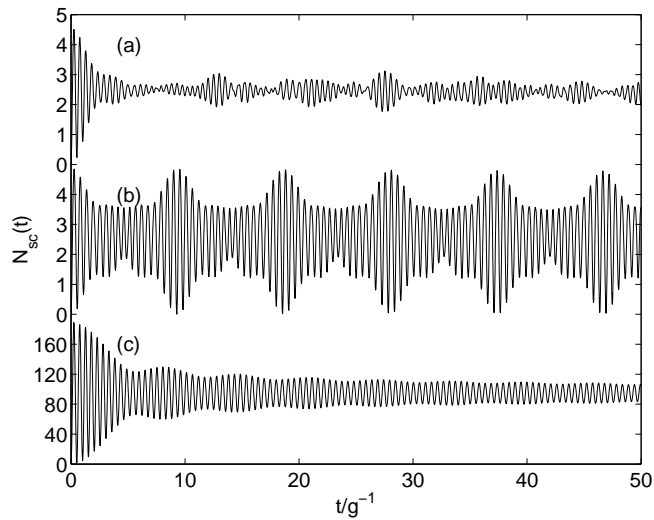}
\caption{Mean number of scattered atoms, $N_{sc}=\sum P_{p_f}$ for
Fock states of the light field and for different recoil
energies.(a) Running wave light field with $N_q=N_{-q}=6$ photons
in each of the two modes with recoil energy $E_{2q}=50g$. (b) Same
light field as in (a) but without dephasing for the atoms,
$E_{2q}=0$. (c) Standing wave light field with $N=6$ photons with
recoil energy $E_{2q}=50g$ for 200 atoms. Time in
units of $g^{-1}$.} \label{fig_rw_fock_mean}
\label{fig_sw_fock_mean}
\end{figure}

Figure \ref{fig_rw_fock_pk}(a) shows $P_{p_f}(t)$ for a Fock state
with $N_q=N_{-q}=6$ photons and five atoms, with a recoil energy
$E_{2q}=50g$ and the Fermi momentum is $k_F=0.1q$. Figure 
\ref{fig_sw_fock_mean}(a) show $N_{sc}(t)$ for the same
parameters.

For short times
the atoms undergo Pendell{\"o}sung oscillations between initial
and final momentum states, with an amplitude that decreases as the
atomic momentum is further detuned from the Bragg resonance
condition. For longer times the oscillations of the individual
atoms dephase, as expected from their different kinetic energies.
However, the dephasing is not as strong as would be the case for a
system of independent particles, such as in the standing-wave case
discussed later on, a clear manifestation of the collective nature
of the system.

In addition, the individual atomic oscillations undergo a decay
that is intrinsically linked to the quantum correlations that
build up between the light field and the atoms and is present even
if we neglect dephasing (i.e. $k_F=0$), as shown in
Fig.\ref{fig_rw_fock_mean}(b). For zero dephasing, the amplitude
of the oscillations eventually revives to its initial value. (The
fact that the collapse of the oscillations and their subsequent
revival resemble a beat phenomenon in the figure is an artifact
from the comparatively small number of atoms and photons.)
Combined with the inhomogeneous dephasing due to the width of the
Fermi sea, this decay results in the total oscillation amplitude
$N_{sc}(t)$ shown in Fig.\ref{fig_rw_fock_mean}(a).

We can gain a qualitative understanding of the collapse and
revival from an analysis of the matrix elements of the operators
$J^+$ and $J^-$, which give an estimate of the transition
frequencies for the atoms from $p_i$ to $p_f$: although for an
initial Fock state the system starts in a state of definite $m$,
it evolves over time into a linear superposition of $m$ states.
The matrix elements of $J^+$ and $J^-$ between different
$m$-states yield different Rabi frequencies and hence Bragg
oscillation periods. Eigenstates of $J^z$ with eigenvalues $m$ and
$m+1$ are coupled by the matrix element
\begin{equation}
\bra{j,m+1}J^+\ket{j,m}=\sqrt{(j+m+1)(j-m)}.
\label{J_matrix_elements}
\end{equation}
We can therefore estimate the collapse time, $T_{\rm decay}$, by
calculating the difference between the fastest and the slowest of
these frequencies, the collapse time being roughly the time after
which this frequency difference has produced a phase difference of
$2\pi$. Under the assumption that all $m$-states contribute
equally to the dynamics we find
\begin{eqnarray}
T_{\rm decay}&=&\frac{2\pi
g^{-1}}{\bra{j,1}J^+\ket{j,0}-\bra{j,j}J^+\ket{j,j-1}}\nonumber\\
&=&\frac{2\pi g^{-1}}{\sqrt{(j+1)j}-\sqrt{2j}}.
\label{t_decay}
\end{eqnarray}

This estimate gives satisfactory agreement with the actual decay
time for small $j$ (it is within $\sim 10\%$ of the numerical
result for our parameters), but breaks down for large $j$. We
attribute this to the fact that the assumption that all $m$ states
are initially equally populated is unphysical for large $j$.

The revival time of the Pendell{\"o}sung oscillations can be
evaluated in a similar fashion: The revivals occur when the Rabi
frequencies for neighboring $m$-states differ in phase by $2\pi$.
This gives
\begin{eqnarray}
T_{\rm revival}&=&\frac{2\pi
g^{-1}}{\bra{j,1}J^+\ket{j,0}-\bra{j,2}J^+\ket{j,1}}\nonumber\\
&=&\frac{2\pi g^{-1}}{\sqrt{(j+1)j}-\sqrt{(j+2)(j-1)}},
\label{trevival}
\end{eqnarray}
which goes to infinity when $j\rightarrow \infty$. Note that this estimate is
not limited to small $j$ values since it does not rest on the
assumption of equal populations of all $m$ states.

\begin{figure}
\includegraphics[width=\columnwidth]{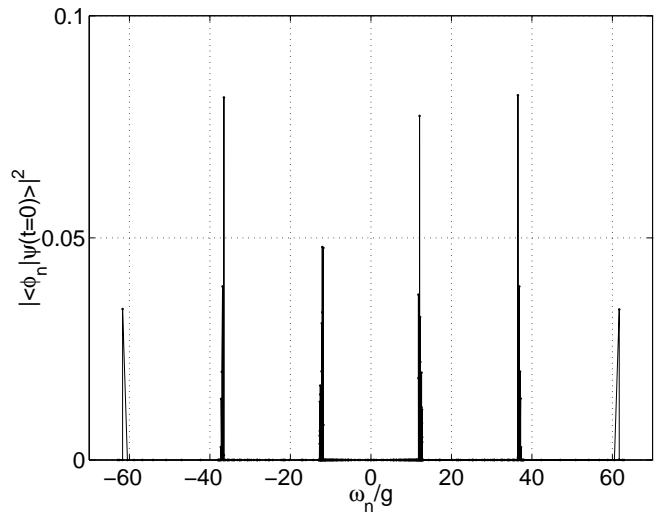}
\caption{Projection of the initial conditions
onto the spectrum of the
Hamiltonian(\ref{bragg_hamiltonian_rw}) for $N_q=N_{-q}=6$ photons,
five atoms, a recoil energy of $E_{2q}=50g$ and a Fermi momentum of
$k_F=0.1q$. Eigenfrequencies in units of $g$.}
\label{projection_example}
\end{figure}

The collapse and revival times can be evaluated more
quantitatively from the spectrum of the Hamiltonian, as
illustrated in Fig. \ref{projection_example}. While the
eigenfrequencies cover the whole spectrum rather densely, the
initial state of the system is well described as a superposition
of just a few groups of eigenstates, and hence only a few narrow
bands of frequencies, which turn out to be almost equally spaced,
significantly determine the atomic dynamics.
If these frequency bands were exactly evenly spaced the
Pendell{\"o}sung oscillations would be perfectly periodic. The
variations in spacing and the widths of the various frequency
bands lead to the more complicated dynamics.

\begin{figure}
\includegraphics[width=\columnwidth]{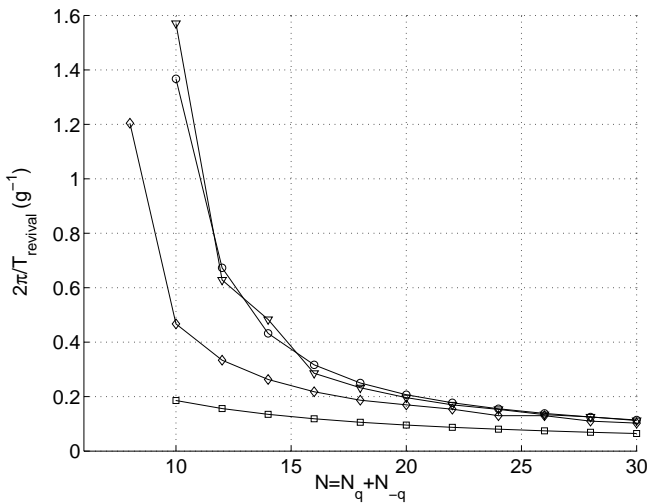}
\caption{Inverse revival time $2\pi/T_{\rm revival}$ of the oscillations
as determined from the spectrum of the Hamiltonian as a function of
the photon number for $E_{2q}=50g$ and $k_F=0.1q$($\Diamond$). For
comparison we also show the results without dephasing, $k_F=0$
($\circ$) and the values obtained from comparing matrix elements
of $J^+$, i.e. $\sqrt{(j+1)j}-\sqrt{(j+2)(j-1)}$ ($\square$). The
inverse revival times as determined directly from simulations for
$N_{sc}(t)$ like the one shown in Fig. \ref{fig_rw_fock_mean}(b)
for no dephasing are also given ($\triangledown$). Frequencies are
in units of $g$.}
\label{fig_scaling_law}
\end{figure}

The width of the frequency bands, which can be traced back to the
usual dephasing of the atoms due to their spread in kinetic
energies, gives the ordinary decay of the density oscillations,
while the variation in separations between bands is a measure of
the inverse revival time. Figure \ref{fig_scaling_law} shows this
separation, obtained numerically for several photon numbers with
and without dephasing. For comparison we also give the inverse
revival time as determined from the matrix elements of $J^+$ as
well as by a direct inspection of $N_{sc}(t)$. While the agreement
between the revival times determined from the spectrum of the
Hamiltonian and from $N_{sc}(t)$ is good for all photon numbers,
the agreement with the estimate based on the matrix elements of
$J^+$ improves for large photon numbers.

\subsubsection{Coherent state}

\begin{figure*}
\includegraphics[width=0.48\textwidth]{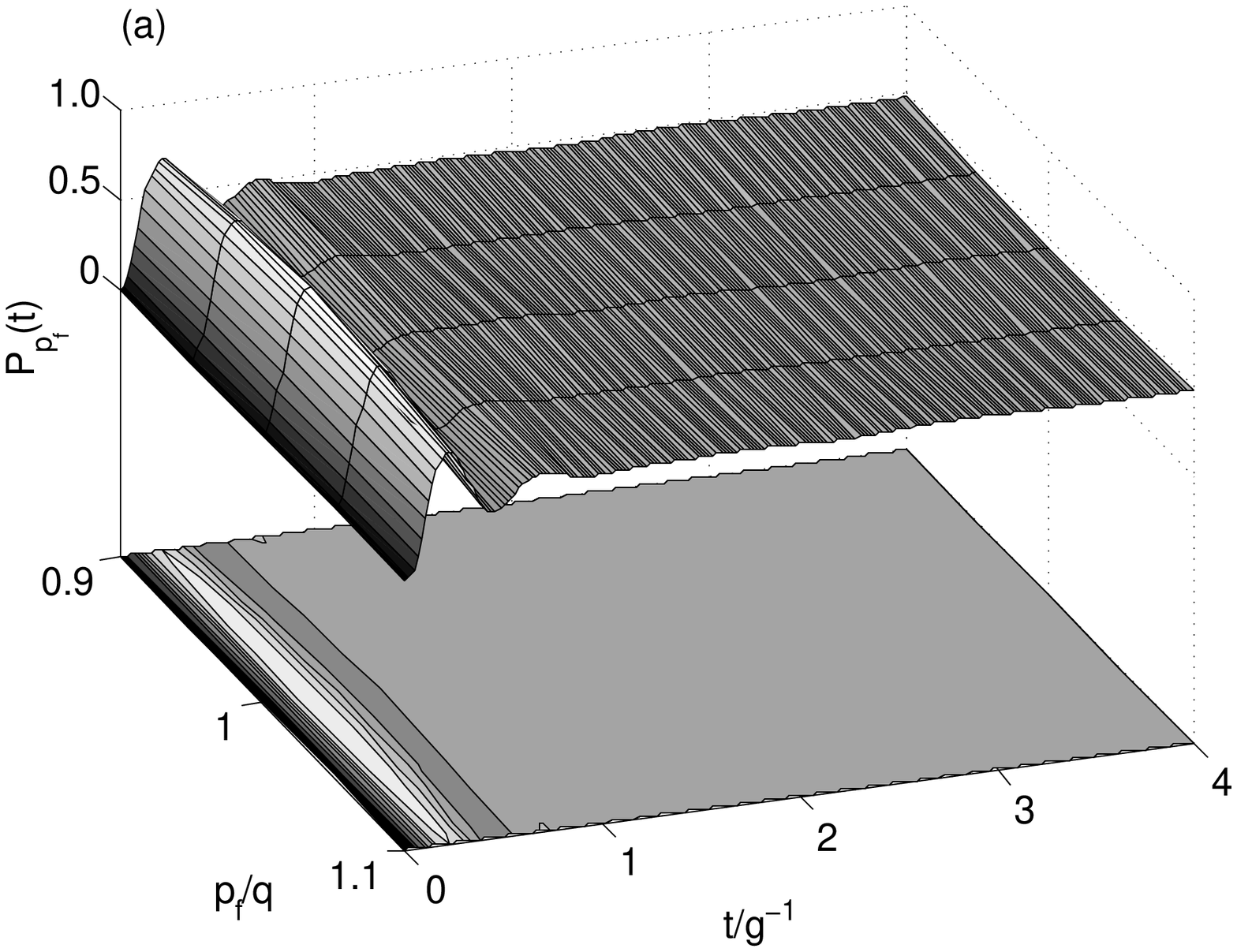}
\includegraphics[width=0.48\textwidth]{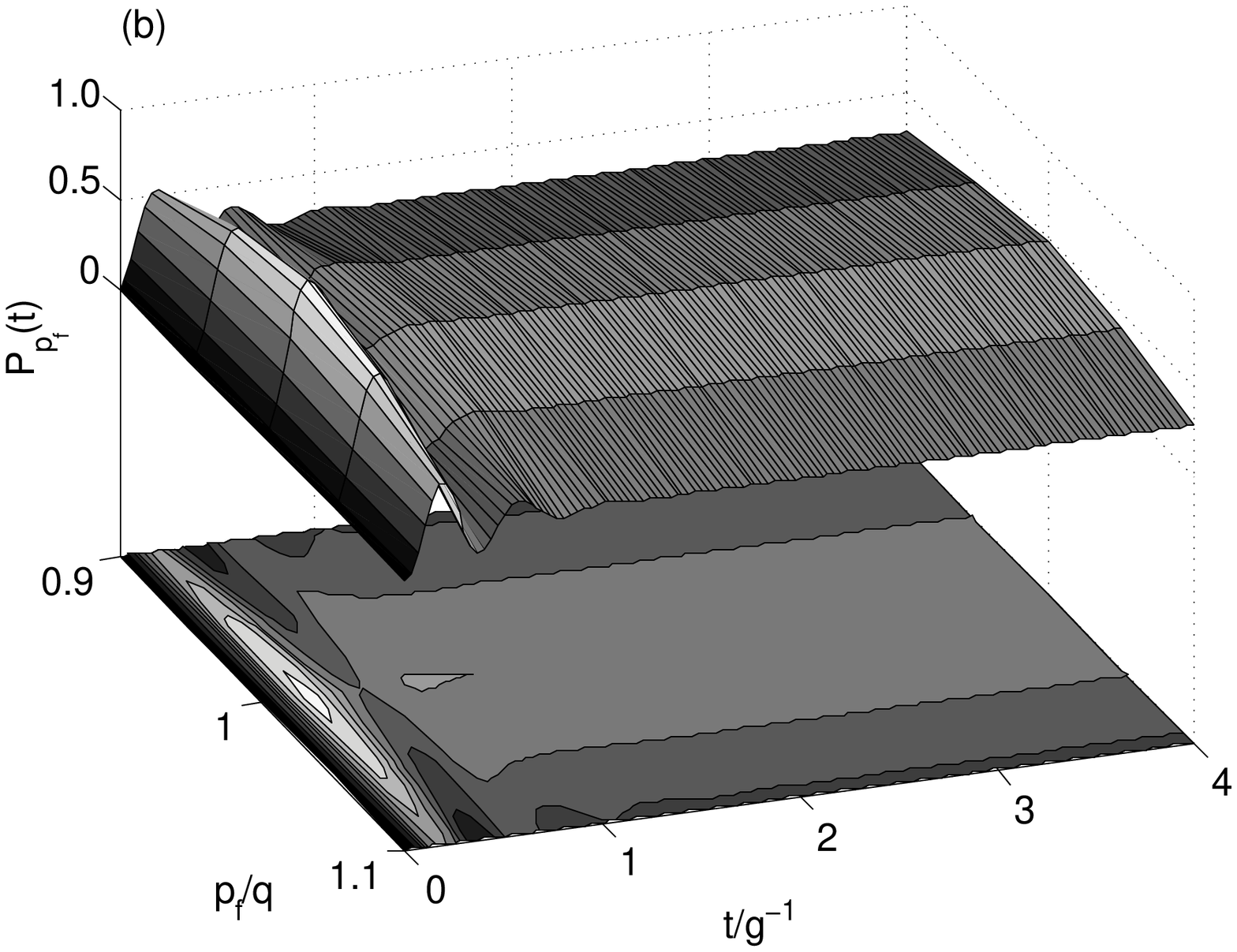}
\caption{Scattering probability for five atoms scattering off (a)
a running wave and (b) a standing wave in a coherent state. In
both cases the recoil energy is $E_{2q}=50g$ and the Fermi
momentum is $k_F=0.1q$. The mean number of photons is
$\overline{N}_q=\overline{N}_{-q}=6$ in (a) and $\overline{N}=12$
in (b). Time is in units of $g^{-1}$ and momentum in units of
$q$.} \label{fig_rw_coherent_pk} \label{fig_sw_coherent_pk}
\end{figure*}

\begin{figure}
\includegraphics[width=\columnwidth]{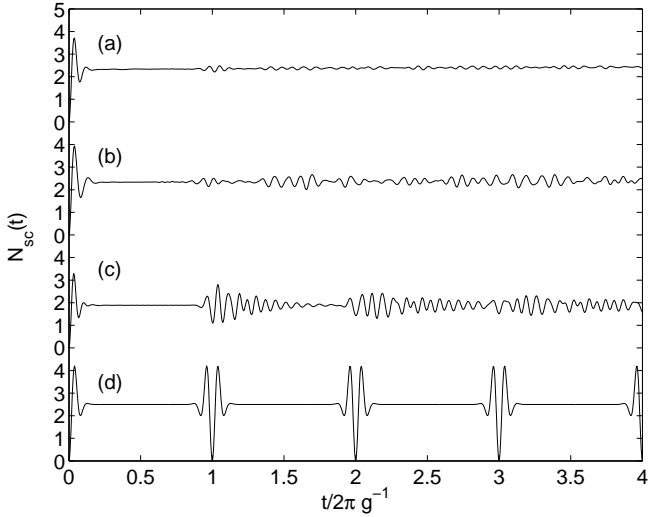}
\caption{Mean number of scattered atoms $N_{sc}(t)$ in coherent
state light fields. (a) Running-wave light field with mean photon
numbers $\overline{N}_q=\overline{N}_{-q}=6$. The Fermi momentum is
$k_F=0.1q$ and the recoil energy is $E_{2q}=50g$. (b) Light field
as in (a), atoms without dephasing, i.e. $k_F=0$. (c)
Standing-wave light field with mean photon number $\overline{N}=12$
and Fermi momentum $k_F=0.1q$ and recoil energy $E_{2q}=50g$. (d)
Light field as in (c), but without dephasing for the atoms,
$k_F=0$. Time in units of $2\pi g^{-1}$.}
\label{fig_rw_coherent_nscattered}
\label{fig_sw_coherent_nscattered}
\end{figure}

The case of a coherent state is readily obtained by averaging the
Fock state results over a Poissonian photon distribution. The
results for $P_{p_f}(t)$ and $N_{sc}(t)$ are given in Figs.
\ref{fig_rw_coherent_pk}(a) and
\ref{fig_sw_coherent_nscattered}(a), respectively. The
oscillations of the mean number of scattered atoms as well as the
Pendell{\"o}sung oscillations of the individual atoms decay in a
time $\lesssim g^{-1}$. In addition to the effects discussed in
the previous section, we now have an additional dephasing due to
the photon statistics of the coherent states. These independent
dephasing processes are normally associated with non-commensurate decay
rates and revival times, hence there are no revivals in this case.

\subsection{Bloch vector model}

If we wish to consider greater atom numbers we can introduce a
factorization scheme such as we used in the Raman-Nath regime.
Here, the fact that the individual atoms are two-level systems in
momentum space suggests that we use instead an approach analogous
to the use of the Bloch vector in conventional quantum optics. We
proceed by introducing the pseudo-spin vector $\vec{S_k}=
S_k^x\uvec{x}+S_k^y\uvec{y}+S_k^z\uvec{z}$
where $\uvec{x}$, $\uvec{y}$, and $\uvec{z}$ are unit vectors
along the $x$, $y$, and $z$ directions in the abstract Bloch
vector space and 
\begin{eqnarray}
S_k^x&=&\frac12(S_k^+ + S_k^-),\nonumber \\
S_k^y&=&\frac{1}{2i}(S_k^+-S_k^-),
\end{eqnarray}
(see Eq. (\ref{pseudospin})). The $S_k^l,l=x,y,z$
obey the usual angular momentum
commutation relations
\begin{equation}
\left[S_k^l,S_{k'}^{m}\right]=i\epsilon_{lmn}\delta_{k,k'}S_k^n,
\label{angular_commutator}
\end{equation}
where $\epsilon_{lmn}$ is the Levi-Civita symbol. Using these
commutation relations we obtain the coupled equations of motion
for the light field and atomic operators
(\ref{bragg_hamiltonian_rw}):
\begin{eqnarray}
\frac{d{\bf J}}{dt}&=&g(S^x \uvec{x} + S^y \uvec{y})\times {\bf J},\nonumber \\
\frac{d{\bf S}_k}{dt}&=&(\delta\omega_k\uvec{z} +g(J^x\uvec{x} +
J^y\uvec{y}))\times \vec{S}_k ,\label{blochequations}
\end{eqnarray}
where we have introduced the
total atomic spin operators
$$
S^l=\sum_k S_k^l. $$ Equations (\ref{blochequations}) are exact
within the two-state approximation of Bragg scattering. We can
obtain a semiclassical picture by factorizing expectation values
of products of atomic and field operators, e.g. $\langle S^m
J^n\rangle = \langle S^m\rangle\langle J^n\rangle$.

For atoms with initial momenta centered around $-q$ we have
$\langle S_k^z(0)\rangle = -1/2$, $\langle S_k^x(0)\rangle = \langle
S_k^y(0)\rangle = 0$ for all $k$, so that the individual atomic Bloch vectors
point to the south pole. Likewise, for a field in a Fock state we
have that $\langle J^x(0) \rangle = \langle J^y(0)\rangle = 0$, so
that ${\bf J}$ points along the $\uvec{z}$-axis, too. From the Bloch
equations (\ref{blochequations}) it is then immediately apparent
that there is no atomic scattering in the semiclassical
description, consistently with the previous discussion.

\begin{figure}
\includegraphics[width=\columnwidth]{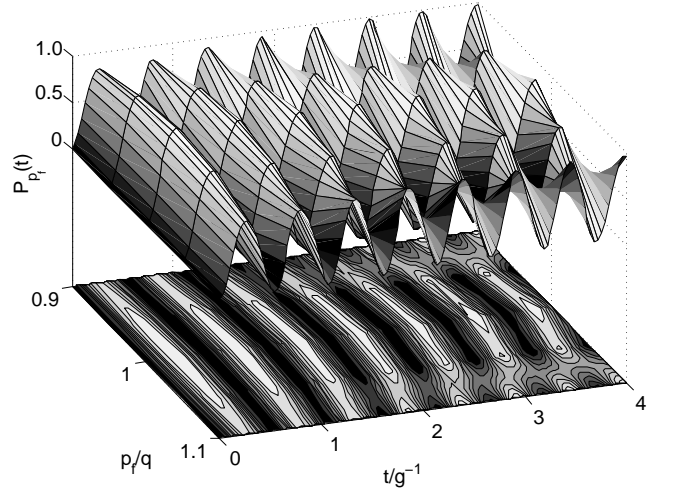}
\caption{Scattering probability $P_{p_f}(t)$ for five atoms in
a running-wave light field as calculated within the Bloch vector
picture. The Fermi momentum is $k_F=0.1q$ and the recoil energy is
$E_{2q}=50g$. Time in units of $g^{-1}$ and momentum in units
of $q$.}
\label{fig_classical}
\end{figure}

For a coherent state, on the other hand, ${\bf J}$ is not parallel
to the $z$-axis. For our choice of phase and for
$\overline{N}_q=\overline{N}_{-q}$, it points instead along the
$x$-direction. The phase relationship between the two
counter-propagating coherent states leads to an intensity grating
and the atoms will scatter off of it. Figure \ref{fig_classical}
shows the resulting scattering probability $P_{p_f}(t)$  obtained
from this approximate model for five atoms and a light field
initially in a coherent state with mean photon numbers
$\overline{N}_q=\overline{N}_{-q}=6$.

The atomic Pendell{\"o}sung oscillations do not decay as fast as
in the full quantum description of Fig.
\ref{fig_rw_coherent_pk}(a). In the present picture, it can be
attributed to the degradation of the intensity grating. The
maximum oscillation amplitude occurs when $\vec{J}$ lies in the
equatorial $\uvec{x}$-$\uvec{y}$ plane, but the scattering of the atoms leads to a
redistribution of the photons between the counter-propagating
modes and a decrease in the optical fringe visibility.

\section{\label{stand_wave}Standing wave quantization}

For a standing-wave quantization
of the light field the Hamiltonian (\ref{basic_hamiltonian_rw}) is
replaced by
\begin{equation}
H_s = \sum_k E_k c^\dagger_k c_k + \frac{g\hat{N}}{2} \sum_k
c^\dagger_{k-q}c_{k+q}
+ H.C.
\label{basic_hamiltonian_sw}
\end{equation}
where $\hat{N}=a^\dagger a$, the number operator for the optical field
mode, is clearly a constant of motion, and $a^\dagger$ and $a$ are
bosonic creation and annihilation operators.

\subsection{Raman-Nath regime}

From Eq. (\ref{basic_hamiltonian_sw}) we now have
\begin{eqnarray}
i\frac{d}{dt}c_{k_1}^\dagger c_{k_2}&=&
\label{eqnofmotion_sw_rn_atoms}
(E_{k_2}-E_{k_1})c_{k_1}^\dagger c_{k_2}\\
&&+\frac{g\hat{N}}{2}\left(c_{k_1}^\dagger c_{k_2-2q} +
c_{k_1}^\dagger c_{k_2+2q}\right)\nonumber\\
&&-\frac{g\hat{N}}{2}\left(c_{k_1+2q}^\dagger c_{k_2}
+c_{k_1-2q}^\dagger c_{k_2}\right).\nonumber
\label{eqnofmotion_sw_rn}
\end{eqnarray}

Since
Eq. (\ref{basic_hamiltonian_sw}) does not couple states with different photon numbers, we can replace 
$\hat{N}$ by the corresponding eigenvalue $N$ for a particular number state, $\ket{N}$. We can then calculate the evolution of 
$\langle c_{k_1}^\dagger c_{k_2}\rangle$ for a general state of the field
by averaging over the appropriate photon number distribution.

The equations for the first-order moments of the individual atoms
are identical to those describing the scattering of a
single atom by a classical light field, if one identifies
$gN/2$ with the classical Rabi frequency. This follows 
from the absence of correlations between the light field
and the atoms, together with the condition
$q>k_F$, which implies the absence of Pauli blocking. 

The scattering of a single atom by a classical field is a
well-studied problem. An analytical solution is known in the Raman-Nath regime, see e.g.
\cite{Rojo:talbot_oscillations,Meystre:elements_of_quantum_optics,
Search:BCS_scattering}. If the kinetic energy of the
atoms is not negligible, on the other hand, one has to rely on a numerical solution. 

\begin{figure}
\includegraphics[width=\columnwidth]{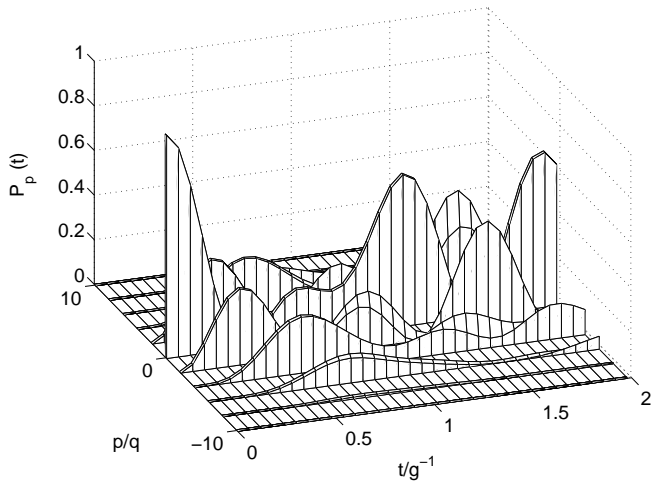}
\caption{Scattering of two atoms off of a standing-wave light field
in a Fock state with $N=6$ photons, the recoil energy is
$E_{2q}=g$ and the Fermi momentum is $k_F=0.1q$. Time 
in units of $g^{-1}$ and momentum in units of photon momentum $q$.}
\label{fig_rn_sw_fock}
\end{figure}

Figure \ref{fig_rn_sw_fock} shows the results for two atoms in
a Fock state with N=6 photons. For short times the
scattering clearly resembles the single-particle behavior,
the probability of finding an atom in the $m$-th side mode
being proportional to $J_m^2(gNt)$. Note that this result is identical to
the approximate first-order calculation for the running-wave coherent state of section III.

\begin{figure}
\includegraphics[width=\columnwidth]{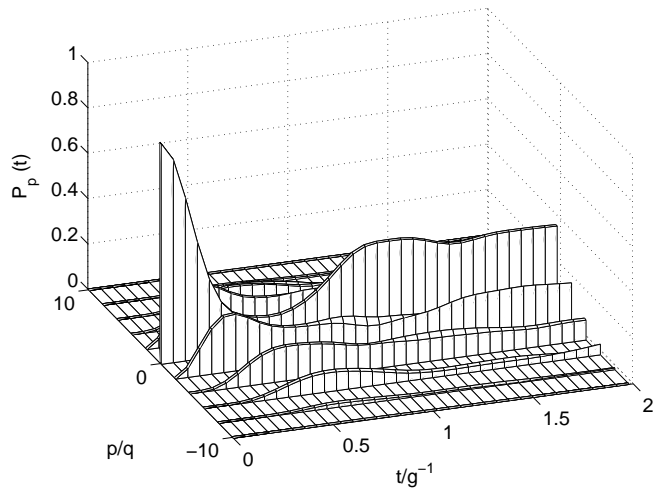}
\caption{Scattering of two atoms off of a standing-wave light field
in a coherent state. The mean number of photons is six, the recoil
energy is $E_{2q}=g$ and the Fermi momentum is $k_F=0.1q$. Time
in units of $g^{-1}$ and momentum in units of photon
momentum $q$.}
\label{fig_rn_sw_coherent}
\end{figure}

The solution for a coherent state is obtained by averaging over
a Poissonian photon number distribution. The
result is shown in Fig. \ref{fig_rn_sw_coherent} for a mean photon number ${\bar N}=6$, which illustrates the dephasing due to the distribution of Rabi frequencies for such a state. 

\subsection{Bragg regime}

As previously discussed, the atoms can now be described in a two-state basis, leading to the effective Hamiltonian, see Eq.
(\ref{bragg_hamiltonian_rw}):
\begin{equation}
H=\sum_{k\in [-k_F,k_F]}
\left\{\delta\omega_k S_k^z +
\frac{g\hat{N}}{2}(S_k^- + S_k^+) \right\}.
\label{bragg_hamiltonian_sw}
\end{equation}
Evidently, the Hamiltonian decomposes into independent single-particle
Hamiltonians for the individual atoms, $H=\sum_k H_k$, so that
the atomic equations of motion decouple and can
be readily solved analytically. If the cavity is in a Fock state, the atoms undergo Pendell{\"o}sung
oscillations independently of each other, with a detuning given by
$\delta\omega_k$. We then obtain

\begin{equation}
P_{p_f}(t)=\frac{(gN/2)^2}{\delta\omega_k^2 +(gN/2)^2}\cdot
\sin^2\left(\sqrt{\delta\omega_k^2+(gN/2)^2} t \right),
\label{ppf}
\end{equation}
as illustrated in Fig. \ref{fig_sw_fock_pk}(b) for five atoms and the light field in a Fock state with $N=12$
photons. As expected, the frequency of the Pendell{\"o}sung oscillations
increases and their amplitude decreases with detuning, see Eq. 
(\ref{ppf}). In contrast to the running-wave case, these oscillations remain perfectly sinusoidal for all times.

The oscillations in the mean number of scattered atoms $N_{sc}(t)$ is shown in
Fig. \ref{fig_sw_fock_mean}(c) for an atom number of
$200$. All other parameters are as before \footnote {This comparatively large number of atoms is to avoid artificial revivals resulting from a finite quantization volume in the numerics, which are the only dephasing effect
in the present case. In all other cases the frequency differences between
the atoms are smaller than all other frequencies in the system
so that the associated revivals show up only at much later times.}.
Their decay due to the spread in
detunings $\delta\omega_k$, resembles the dephasing
of an inhomogeneously broadened ensemble of independent two-level
systems, with a dephasing time given
by $\frac{2\pi}{\delta\omega_{k_F}}$.

The scattering probability for a coherent state light
field is shown in Fig. \ref{fig_sw_coherent_pk}(b) and
the mean number of scattered atoms $N_{sc}(t)$ is shown in
Fig. \ref{fig_sw_coherent_nscattered}(c). The atomic parameters
are the same as for the Fock state calculations while
the light field has been replaced by a coherent state
with the mean number of photons $\overline{N}=12$.

As in the running-wave case, the oscillations collapse due to the
spread in Rabi frequencies associated with the coherent state. In
contrast to that former case, though, they also undergo a revival
after the time $T_{\rm revival}=2\pi g^{-1}$, as 
follows from the fact the difference
between the Rabi frequencies for $N$ and $N+1$
photons is $g/2$, independently of $N$. As in the two-photon Jaynes-Cumming model, all number
states rephase at the same instant and the revivals are perfect if
the dephasing due to the kinetic energy of the atoms can be neglected, as shown in
Fig.\ref{fig_sw_coherent_nscattered}(d). With dephasing included, the revivals
are only partial, see Fig. \ref{fig_sw_coherent_nscattered}(c).

Another difference between the running-wave and standing-wave quantization schemes can be seen in the evolution of the scattering probability
for the coherent state. In the case of running waves the probability of finding an atom in the scattered states is about 1/2, independently of their detuning, while in the standing wave case
this probability decreases with increasing detuning. This difference
underlines the importance of the correlations between the light field
and the atoms. While the atoms move independently in a standing
wave light field, the atoms and the field
become an inseparable quantum system in the running-wave case.

\section{\label{discussion}Conclusion and outlook}

In this paper we have compared the scattering of ultra-cold fermions by
quantized light fields composed of ``true'' standing waves and of superpositions of counterpropagating running waves, both in the Raman-Nath
and in the Bragg regime. The central difference between the two quantization schemes  is that the entanglement between the light field and the atoms plays a crucial role for a
running-wave light case, but not for standing-wave quantization.

In the Raman-Nath regime the scattering by a standing-wave light
field in a Fock state is similar to the scattering by a
classical field, and can be largely understood from the results for a
single atom. The only many-particle effect is
a dephasing of the atomic motion due to the finite width of the
initial momentum distribution of the atoms. For running-wave quantization, on the other hand, the quantum correlations that develop between the light
field and atoms are essential, and their neglect leads to the absence of atomic diffraction by a Fock state of the field. 

In the Bragg regime and for a standing-wave light field, atomic diffraction
can be solved in terms
of solutions for single atoms, which undergo Pendell{\"o}sung oscillations that undergo a
series of collapse and revivals in the case of a coherent light field. For the case of running-wave quantization, 
these oscillations decay even for
a Fock state, a consequence of the correlations that develop between the light
field and the atoms. A
coherent state merely leads to a faster collapse. 

Table \ref{time_scales_table} gives a summary of the
mechanisms that lead
to collapses and possibly revivals in the various cases that we have disccussed, as well as the associated time scales.

\begin{table*}
\begin{ruledtabular}
\begin{tabular}{|p{2.5cm}|c|c|c|c|p{2.250cm}|p{3.25cm}|}
Decay mechanism&\multicolumn{4}{c|}{Present in}&Collapse time&Revival
time\\
&Fock state,&Fock state,&Coherent state,&Coherent state&&\\
&standing wave&running wave&standing wave&running wave&&\\
\hline
Dephasing due to kinetic energy of atoms&yes&yes&yes&yes&
$2\pi\frac{q}{2k_F E_{2q}}$&
---\\
\hline
Different Rabi frequencies due to photon number
uncertainty&no&no&yes&yes&
$4\pi/\sqrt{\overline{N}}g$&
$2\pi/g$\\
\hline
Correlations between light field and atoms&no&yes&no&yes&
$\frac{4\pi}{g\left(\sqrt{(N+2)N}-2\sqrt{N}\right)}$&
$\frac{4\pi}{g\left(\sqrt{(N+2)N}-\sqrt{(N+4)(N-2)}\right)}$
\end{tabular}
\end{ruledtabular}
\caption{Time scales for the different processes that lead to a
collapse and revival of the Bragg oscillations.}
\label{time_scales_table}
\end{table*}

This paper has only considered fermionic atoms. Bosonic operators commute rather than
anticommute, resulting in different equations of motion
for the second- and higher correlation functions of atomic operators.
Consequently, differences in the scattering result solely
from the effect of these higher-order correlations on the dynamics.

This work is supported in part by the US Office of Naval Research, by
the National Science Foundation, by the US Army Research Office, by
the National Aeronautics and Space Administration, and by the Joint
Services Optics Program. One of the authors (D. M.) is supported
by the DAAD of Germany. We would like to thank T. Miyakawa and
M. J\"a\"askel\"ainen for fruitful discussions.

\bibliography{draft10}

\end{document}